\documentclass[preprint]{aastex}

\shorttitle{X-rays in PMS Stars}
\shortauthors{Stassun et al.}

\begin{document}
\slugcomment{To appear in AJ}

\title{X-ray Properties of Pre--Main-Sequence Stars 
in the Orion Nebula Cluster with Known Rotation Periods}

\author{Keivan G.\ Stassun\altaffilmark{2},
David R.\ Ardila\altaffilmark{3},
Mary Barsony\altaffilmark{4},
Gibor Basri\altaffilmark{5},
and
Robert D.\ Mathieu\altaffilmark{6}}

\altaffiltext{2}{Department of Physics \& Astronomy,
Vanderbilt University, Nashville, TN 37235; keivan.stassun@vanderbilt.edu}
\altaffiltext{3}{Johns Hopkins University,
Baltimore, MD }
\altaffiltext{4}{San Francisco State University}
\altaffiltext{5}{University of California, Berkeley}
\altaffiltext{6}{University of Wisconsin--Madison}

\begin{abstract}
We re-analyze all archival {\it Chandra/ACIS} observations of the Orion Nebula 
Cluster (ONC) to study the X-ray properties of a large sample of 
pre--main-sequence (PMS) stars with optically determined rotation periods. 
Our goal is to elucidate the origins of X-rays in PMS stars
by seeking out connections between the X-rays and the mechanisms 
most likely driving their production---rotation and accretion. 
Stars in our sample have $L_X / L_{bol}$ near, but below, the 
``saturation" value of $10^{-3}$.  In addition, in this sample
X-ray luminosity is significantly correlated with stellar rotation,
in the sense of {\it decreasing} $L_X / L_{bol}$ with more rapid rotation. 
These findings suggest that stars with optical rotation periods are 
in the ``super-saturated" regime of the rotation-activity relationship, 
consistent with their Rossby numbers. 
However, we also find that stars with optical rotation periods are
significantly biased to high $L_X$. 
This is not the result of magnitude bias in the
optical rotation-period sample but rather to the diminishingly small amplitude 
of optical variations in stars with low $L_X$. Evidently, there exists in 
the ONC a population of stars whose rotation periods are
unknown and that possess lower average X-ray luminosities than 
those of stars with known rotation periods. 
These stars may sample the linear regime of the rotation-activity relationship. 
Accretion also manifests itself in X-rays, though in a somewhat 
counterintuitive fashion: While stars with spectroscopic signatures of accretion
show {\it harder} X-ray spectra than non-accretors, they show {\it lower} 
X-ray luminosities and no enhancement of X-ray variability. 
We interpret these findings in terms of a common origin for the X-ray emission
observed from both accreting and non-accreting stars, with the X-rays from
accreting stars simply being attenuated by magnetospheric accretion columns. 
This suggests that X-rays from PMS stars have their origins 
primarily in chromospheres, not accretion.
\end{abstract}


\section{Introduction\label{intro}}
X-rays serve as one of our primary probes of magnetic activity in
solar-type stars. 
Thus, much of our understanding of
key physical processes thought to be connected to magnetic fields---stellar 
winds, for example---derives from the study of stellar X-rays.

One of the most compelling stories in stellar astrophysics told through 
X-rays is that of the intimate relationship between stellar rotation and magnetic
field generation. Indeed, among late-type main-sequence stars, rotation
is the strongest correlate of X-ray luminosity, and the observed
rotation/X-ray relationship 
\citep{pallavicini81,caillault96,randich97,jeffries99,randich00} 
has become central to the current paradigm of dynamo-generated 
magnetic fields, of magnetically driven stellar winds, and 
ultimately of the evolution of stellar angular momentum.

The observed relationship between rotation and X-ray emission on
the main sequence is remarkably clean, and clearly separates stars 
into three regimes (cf.\ \citet{randich00}) 
typically described phenomenologically as the linear, saturated, and 
super-saturated regimes, in order of increasing stellar rotation \citep{randich97}. 
For slowly rotating stars the X-ray luminosity, $\log L_X$, 
scales linearly with the stellar angular velocity, $\log\Omega$,
consistent with the theoretical idea that more rapid stellar rotation 
produces a stronger magnetic field through an $\alpha - \Omega$-type 
dynamo (for stars with radiative cores and convective envelopes) or
through a distributed turbulent dynamo (for fully convective stars). 
For stars rotating more rapidly than a certain threshold, the X-ray
luminosity is observed to ``saturate" at a fixed value relative to the
stellar bolometric luminosity: $\log L_X / L_{bol} \approx -3$. While the reasons
for saturation have not been well understood, this observation has had
important ramifications for efforts to model the angular momentum evolution 
of young solar-type stars \citep{krishnamurthi97,bouvier97}. The models
now routinely include saturation as a key ingredient in their 
parametrizations of angular momentum evolution through winds.
Finally, the most rapidly rotating stars exhibit X-ray emission at levels
roughly a factor of 2 below the saturation value. This
``super-saturation" effect \citep{james00} has been poorly understood.
\citet{barnes03a,barnes03b} has recently re-interpreted 
super-saturation in terms of a new paradigm
for the angular momentum evolution of solar-type stars;
whether this new paradigm will survive detailed scrutiny is not yet clear. 

What is clear is that the rotation/X-ray 
relationship serves as a key observational touchstone for developing
and evaluating our theoretical understanding of the generation
and evolution of stellar magnetic fields, the generation and evolution 
of stellar winds, and the evolution of stellar angular momentum. 
Considerable observational effort has been invested, therefore, in 
trying to establish the presence of a rotation/X-ray relationship among
pre--main-sequence (PMS) stars, where the questions of magnetic field
generation and evolution, winds, and angular momentum evolution remain
largely unanswered. Unfortunately, to date these efforts have not borne 
much fruit. 
In one recent study of T~Tauri stars (TTS) in Taurus-Auriga
\citep{stelzer01}, a rotation/X-ray correlation has been
reported, but the sample size is small ($N=39$) and there are lingering concerns
with respect
to completeness/reliability of the rotation periods and biases in the
sample both astrophysical and observational. In particular, as discussed by 
\citet{feig03},
the Taurus-Auriga PMS population appears to be deficient in both high-mass stars 
and faint, low-mass weak-lined TTS. In addition, rotation periods in
this region were determined from photometric monitoring campaigns
some 10 years ago that were relatively sparsely sampled and of relatively
short duration, and therefore potentially biased against both very fast
and slow rotators.

Two new studies based on deep {\it Chandra} observations of the 
Orion Nebula Cluster (ONC) provide the most comprehensive
analyses yet of X-rays and rotation in a large, coeval ($\sim 1$ Myr)
sample of
PMS stars. Neither the study of Flaccomio and collaborators
\citep{flac-basic,flac-acc,flac-time} with the High Resolution Camera (HRC), 
nor that of Feigelson and collaborators \citep{feig02,feig03} with the 
Advanced CCD Imaging Spectrometer (ACIS),
found evidence for a rotation/X-ray relationship such as
that observed on the main sequence or that reported
for Tau-Aur TTS by \citet{stelzer01}. 
Indeed, these studies find that it is stellar mass that is by far
the dominant correlate of PMS X-ray luminosity, with
$\log L_X/L_{bol}$ correlating with rotation either not at all, or perhaps 
slightly in the opposite sense from the main-sequence rotation/X-ray 
relationship. 

Both studies found that stars with known rotation periods have
X-ray luminosities near the main-sequence saturation value of 
$\log L_X / L_{bol} \approx -3$. 
Moreover, both studies suggest that PMS stars might in fact be 
{\it expected} to reside in the super-saturated regime, considering that 
typical PMS Rossby numbers \citep{kim-demarque,ventura}
are small due to the long convective turnover times 
($\tau_c \sim 800$ days) of these very young and fully convective
stars. Thus, while direct observation of PMS stars in the linear regime
of the rotation/X-ray relationship 
remains elusive, these studies seem to confirm, if
indirectly, the basic picture of the rotation/X-ray relationship by 
suggesting that all ONC stars are in the super-saturated regime,
and that this is where they ought to be.

But not all ONC stars detected by {\it Chandra}
have known rotation periods. To what extent is the sample with 
periods representative of the entire ONC population in terms
of X-ray properties? If they are not representative, how does this
group differ in other salient characteristics, such as accretion,
and how might these differences affect our interpretation of the
origin of X-rays in PMS stars?
In addition, while it is clear from previous analyses that X-ray 
emission from PMS stars is not temporally static---X-ray flaring is 
ubiquitous in the ONC---it is not yet clear whether or to what extent 
X-ray variability may be affecting our ability to measure reliable
X-ray luminosities.
Might X-ray flaring be scrambling the signal of an underlying 
rotation/X-ray relationship? 
That some stars flare while others do not
is interesting in its own right: Do the flaring characteristics of stars
with known rotation periods represent those of all stars?  Again, how
might differences here affect our interpretation of the origin of PMS 
X-rays?

Motivated by these questions, we have re-analyzed all archival 
{\it Chandra}/ACIS observations of the ONC that include stars with known
rotation periods. Our aim is to derive X-ray luminosities for as large
a sample of known rotators as possible, employing a consistent analysis
scheme throughout, including filtering of flares in the hopes of
minimizing the effects of X-ray variability.

In \S 2 we describe the data used and our processing/analysis
procedures. We then present our basic results in \S 3, first focusing 
on the X-ray nature of the rotator sample as compared to the entire
ONC population. We show that stars with known rotation periods are
significantly more X-ray luminous, and more likely to be X-ray variable, 
than stars for 
which rotation periods have not been measured. We then explore the relationship
between X-rays and rotation. We find that most stars with known rotation
periods appear to be in the super-saturated regime, having 
$\log L_X / L_{bol} \lesssim -3$, with a statistically significant correlation
in which faster rotators have lower X-ray luminosities.
But we also find that stars without rotation periods---being 
less X-ray luminous on average---show a range of $L_X/L_{bol}$ comparable
to that observed on the main sequence. These stars may represent the 
beginnings of the linear regime of the
rotation/X-ray relationship. Finally, we explore the
relationship between X-rays and accretion. We find
that while stars with spectroscopic signatures of accretion show harder 
X-ray spectra than non-accretors, they also show lower X-ray luminosities and no
enhancement of X-ray flaring. 

We discuss the implications of our findings in \S 4, where we (a) emphasize that 
current rotation-period measurements in the ONC have not probed the full range
of underlying stellar X-ray properties, (b) suggest that a main-sequence type
relationship between X-rays and rotation may in fact be present in the ONC, 
and (c) argue that the data
imply a chromospheric---not accretion---origin for X-rays from
PMS stars.  We summarize our conclusions in \S 5.

\section{Data\label{data}}
Our primary goal is to study the relationship of stellar X-rays to stellar 
rotation among a large sample of PMS stars in the ONC. Key parameters in
our analysis are the stellar rotation period and the ratio of the X-ray
luminosity, $L_X$, to the bolometric luminosity, $L_{bol}$.
We restrict our analysis to observations with ACIS because its energy
resolution allows $L_X$ to be determined from fits to the X-ray spectral 
energy distribution. We also consider only reasonably long observations 
($\sim 100$ ksec) so that we can attempt to derive quiescent X-ray
luminosities by filtering out flaring events.

Thus our study sample comprises ONC stars that:
(1) have known rotation periods, (2) have derived bolometric luminosities,
and (3) have been observed by the {\it Chandra} ACIS instrument.
Where possible we would also like to study the relationship of X-rays to
accretion, so we include such measurements where available.

Here we describe the data from the literature that we compile to form our
study sample (\S \ref{literature}). We also describe the data from the 
{\it Chandra} archive that we use (\S \ref{chandra}) as well as the 
procedures employed in their reduction (\S \ref{reduction}) and analysis 
(\S \ref{analysis}). We close with a brief discussion of our
assessment of the quality and reliability of the X-ray measurements 
(\S \ref{quality}).

\subsection{Supporting data from the literature\label{literature}}
Rotation period measurements are available 
from the optical studies of \citet{stass99} and \citet{herbst02} 
for 431 PMS stars in the ONC
which were also included in the optical photometric/spectroscopic study
of \citet{hill97}. The latter study provides bolometric luminosities and 
other basic stellar parameters (i.e.\ masses, effective temperatures, 
extinctions, etc.) for most (358) of these stars. In addition, the 
study of \citet{hill98} provides spectroscopic measures of accretion in
the form of \ion{Ca}{2} equivalent widths. 
In Table \ref{table-sample} we summarize our study sample, comprising 220
unique stars with rotation periods that we detect in the {\it Chandra} 
observations described
below. We include relevant stellar properties taken from the sources above. 

\subsection{{\it Chandra} archival data\label{chandra}}
There are three ONC observations in the {\it Chandra} archive relevant to 
this study, two obtained by Garmire (Obs.\ ID's 18 and 1522)
and one by Tsujimoto (Obs.\ ID 634). The Garmire observations are described
by \citet{feig02} and include a 45.3 ksec exposure obtained on 1999 Oct 12--13
and a 37.5 ksec exposure obtained on 2000 Apr 1--2, both centered on the
Trapezium. The Tsujimoto observation, described in \citet{tsujimoto02}, is a
single 89.2 ksec exposure centered on the OMC-2/3 region (just North of the 
Trapezium) obtained on 2000 Jan 1--2.

In all three exposures, the four ACIS-I chips were operational with a 
total field of view of $17\times 17$ arcmin. In addition, all three
exposures had the ACIS-S2 chip in
operation, which is separated from ACIS-I by 2.7 arcmin and has a
field of view of 8.3 arcmin. Finally, the second Garmire exposure 
included the ACIS-S3 chip, again with a field of view of 8.3 arcmin.
We include the ACIS-S data here for completeness, but note that this
results in only a few additional sources due to the highly degraded
point spread function (PSF) of the instrument at large off-axis angles.
The ACIS instrument measures photon arrival times, positions, and energies
(0.5--8 keV),
so that for each detected source an X-ray light curve and spectral energy
distribution can be constructed.

\subsubsection{Reduction\label{reduction}}
We reprocessed all three exposures in the same manner, starting from
the {\tt \_evt1} event files\footnote{Event files consist of 
arrival times, positions, energies, and other information for each detected
X-ray photon.}, using the standard
CIAO\footnote{Chandra Interactive Analysis of Observations (CIAO) version 2.2.} 
procedure {\tt process\_acis\_events}
and updated calibration files obtained from the {\it Chandra} X-ray
Center\footnote{See \url{http://cxc.harvard.edu}.} (CXC) in Sept 2002. Photon 
events were filtered according to their grade and status flags, and the images
destreaked, following the standard CXC science threads. We also manually
updated the astrometric header keywords based on the latest astrometric
calibration available from the CXC.

The resulting event files ({\tt \_evt2} files) were then searched for
point sources using the CIAO task {\tt celldetect}. The task uses a spatially
variable PSF, and we kept only those 
sources with a signal-to-noise ratio (SNR) of 5 or greater. We set the
{\tt celldetect} task to return source ellipses with a size of 99\%
encircled energy, and defined a background annulus whose inner and outer 
semi-major axes were, respectively, 1.5 and 1.7 times
larger than the source ellipse.

To make the photon extraction computationally feasible, at each source 
position we then extracted a sub-region just larger than the background
ellipse, using a set of IDL\footnote{Interactive Data Language} scripts
developed by us. Thus for each of the three exposures, the result 
of the reduction step is a set of event files, one for each of the 
detected sources.

\subsubsection{Analysis\label{analysis}}
With a set of event files corresponding to each source detected with SNR $>5$,
we next applied an automated time-filtering of each source light curve in 
order to remove flare events prior to modeling the X-ray spectral energy
distribution (SED) to derive X-ray luminosities. The aim of this procedure 
is to determine a quiescent $L_X$ for each source. Based on the documented
sensitivity limits of ACIS, in all that follows we
use only X-ray photons with energies in the range 0.5--8 keV.

\begin{figure}
\epsscale{0.85}
\plotone{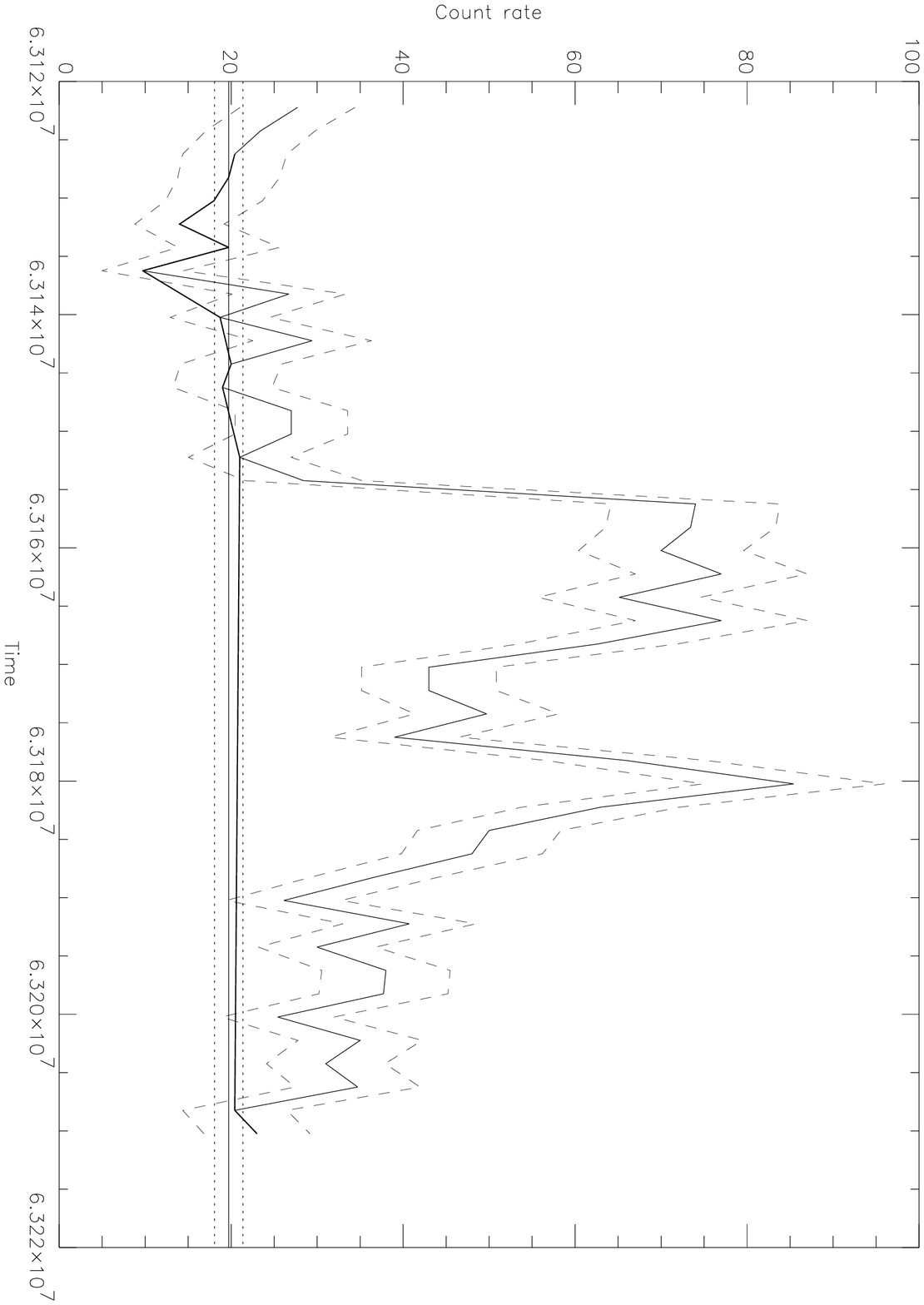}
\caption{Example of light curve filtering for flare events for a source
in the observation of Tsujimoto. The thin solid line represents the 
observed light curve, and dashed lines represent $1\sigma$ errors
based on simple counting statistics. The thick solid line represents
the light curve after flare filtering. The horizontal solid and dotted
lines indicate the quiescent count rate determined from the filtering
procedure (solid line) and $1\sigma$ errors (dotted).
\label{fig-lcfilt}}
\end{figure}

The time-filtering of the light curves was implemented in IDL using 
procedures developed by us. For each source, the process involves
the following steps (see example in Fig.\ \ref{fig-lcfilt}):
(1) Construct source and background light curves using the CIAO
{\tt lightcurve} script with a binning interval of 2 ksec;
(2) subtract background light curve from source light curve;
(3) exclude bins that are $>3\sigma$ brighter than the median, which is
computed from the lowest 15\% of the bins;
(4) re-determine the median and again exclude deviant bins, iterating
until no more bins are excluded; and
(5) output a new event file that includes only the time intervals
of the surviving bins.

With time-filtered event files in hand for each of the detected sources,
we determined the $L_X$ of each source via a standard spectral analysis
using {\tt SHERPA}. For each source in each of the three exposures, the 
position-dependent auxiliary response file (ARF) and redistribution
matrix file (RMF)\footnote{The ARF contains the combined 
telescope/filter/detector areas and efficiencies as a function of energy.
The RMF translates detector pulse heights into photon energies.} 
were computed with the CIAO {\tt psextract} command and a model spectrum was
fit. The model used was a two-component thin thermal plasma with 
absorption by an intervening column of hydrogen. The free parameters of
the model are the absorbing hydrogen column density ($\log N_H$), the 
temperatures of the two plasma components ($kT_1$, $kT_2$), the 
metallicity ($Z$), and a normalization (scaling) coefficient for each 
plasma component. A $\chi^2$ minimization procedure was used to fit 
each source's SED for these parameters, iterating until convergence
was achieved.

Given the large number of free parameters, there is no guarantee
that the best fit adopted is truly a global best fit or even that there
is only one possible global best fit. Thus we emphasize that our goal in
the spectral fitting is not the values of the model parameters themselves; 
we simply seek 
a reasonably good fit from which we can determine the X-ray luminosity
of the source. The fit can thus be thought of as a (possibly over-determined)
spline fit to the X-ray SED of the source, which we then integrate to
measure the source flux, $F_X$. Adopting a distance of 470 pc to the
ONC we convert the measured $F_X$ values into $L_X$.

In principle, we can correct each $L_X$ for intervening absorption
using the value of $\log N_H$ determined from the spectral fit. However,
\citet{feig02} have demonstrated that the $\log N_H$ values determined
from spectral fitting do not correspond very well to $A_V$ values
determined from optical photometry/spectroscopy. We thus follow
\citet{feig02} and choose not to correct the measured $L_X$ for
absorption.

With $L_X$ values determined for each source from each of the three
{\it Chandra} exposures, we match the sources with known rotation periods
by searching for a positional match within
the error ellipses of the detected X-ray sources. 
We find 220 stars with rotation periods in the {\it Chandra} images.
In cases where a
given target is detected in more than one {\it Chandra} exposure, we
select for our subsequent analysis the {\it lower} value of $L_X$,
assuming that the source changed its intrinsic $L_X$ between 
observations, and that the lower value represents the best estimate of
the quiescent $L_X$. 

The X-ray properties of these 220 sources 
(Table \ref{table-sample}) are summarized
in Table \ref{table-lx}, which includes all $L_X$ measurements of each 
source (as many as three different measurements because there are three
separate exposures). In addition, Table \ref{table-lx} provides the
$L_X$ measurements from \citet{feig02} for comparison\footnote{We do 
not include X-ray luminosities derived by \citet{tsujimoto02} for 
comparison as their tabulated values are corrected for extinction, 
while those reported here and by \citet{feig02} are not.}.
Those authors detected 253 stars\footnote{The \citet{feig02} study detected a
total of 1075 stars. Included in the {\it ACIS} field of view were 263
stars with rotation periods; 10 stars with rotation periods were not
detected by \citet{feig02}.}
with rotation periods, and here we
re-detect 190 of them, presumably due to our higher SNR criterion ($>5$)
for source detection (see above). 
The 30 stars\footnote{Included in the {\it ACIS} field of view were 35 stars 
with rotation periods; 5 stars with rotation periods were not detected by us.}
with rotation periods detected by us and not by 
\citet{feig02} (see Table \ref{table-lx}) derive from the \citet{tsujimoto02} 
exposure.

Table \ref{table-lx} also includes a
descriptor for the variability of each source's light curve. These are
taken from \citet{feig02} when the source was included in that study;
otherwise, the descriptor is assigned by us following the procedure
of \citet{feig02}. A designation of `Const' indicates that the light curve
shows an approximately constant flux with time; `LTVar' indicates 
statistically significant variability that occurs slowly in time, 
resulting in a different mean flux level in the different ACIS exposures;
`Flare' indicates a statistically significant variation on short timescales;
and `PosFl' indicates a flare-like variation of marginal significance.
Finally, Table \ref{table-lx} provides a subjective
quality flag for each $L_X$ determination, which we now discuss.

\subsubsection{Quality assessment\label{quality}}
Since our re-reduction of the archival {\it Chandra} data used updated 
calibrations, and because 
our analysis procedures included time-filtering of flares that other
authors have not done, in this section we assess the reliability of
our reductions. We begin by comparing the $L_X$ values derived
by us to those derived by other authors for the same sources. We
then discuss some specific cases in detail in order to illustrate the
vagaries inherent to this type of analysis.

To start, we visually inspected the {\tt SHERPA} fit of each source
and subjectively flagged those sources whose $L_X$ values we deemed
unreliable due either to an observed spectrum with few counts or to
an otherwise poor fit. The result of this procedure is 154 sources whose
spectra and corresponding spectral fits we felt were subjectively
reasonable. We restrict all subsequent discussion to these 154 sources,
which are indicated in Table \ref{table-lx} by a quality flag of `1'.

\begin{figure}
\plotone{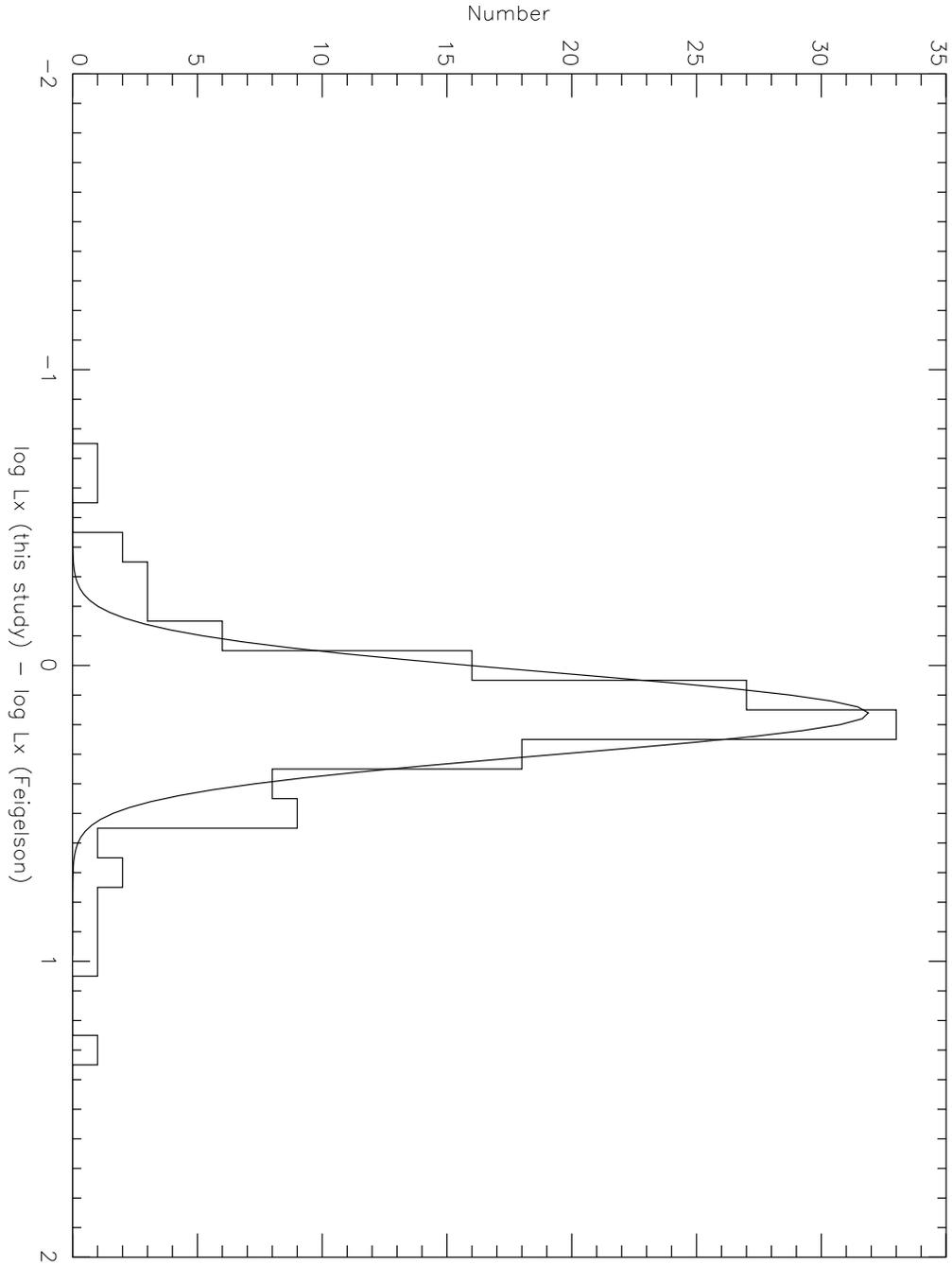}
\caption{Differences between the $L_X$ values measured by us and those
reported by \citet{feig02} (histogram). The gaussian fit shown has
$\sigma = 0.14$ dex and an offset of 0.15 dex. Approximately 0.04 dex
of this offset is due to the different distances assumed to the ONC
by us (470 pc) and by \citet{feig02} (450 pc). 
\label{fig-lxcomp}}
\end{figure}

In Fig.\ \ref{fig-lxcomp} we compare the $L_X$ values obtained by us
to those obtained by \citet{feig02} for the common sources. We basically 
find good agreement between the two sets of measurements. A gaussian
fit to the differences between the two measurements results in a 
standard deviation of $\sigma = 0.14$ dex, an offset of 0.15 dex
(our measurements being systematically larger), and a small number of
outliers. 

Approximately 0.04 dex of
the systematic offset can be accounted for by the fact that we assume
a distance to the ONC of 470 pc while \citet{feig02} assume a slightly
lower value of 450 pc. The remaining difference of 0.1 dex remains unaccounted 
for, but is not surprising given small differences in the calibrations used
in our data reprocessing. On the whole, then, we can report reproducibility 
of the derived $L_X$ to a level of $\sim 0.1$ dex, despite differences in 
calibration, our time-filtering of flares from the light curves, and so on.

Nonetheless, a few stars have very different $L_X$ measurements from the
two analyses (up to about 1 dex). As an example, we consider
star 116, which is the most discrepant between our measurements
and that reported by \citet{feig02}. From the two Garmire exposures we
measure $L_X$ values for this source of $10^{30.7}$ erg/s and 
$10^{30.4}$ erg/s, which encouragingly are similar to one another, but 
are very different from the \citet{feig02}
value of $10^{29.1}$ erg/s (see Table \ref{table-lx}). This is a 
remarkable difference considering that these values derive from 
the same photons.

\begin{figure}
\plotone{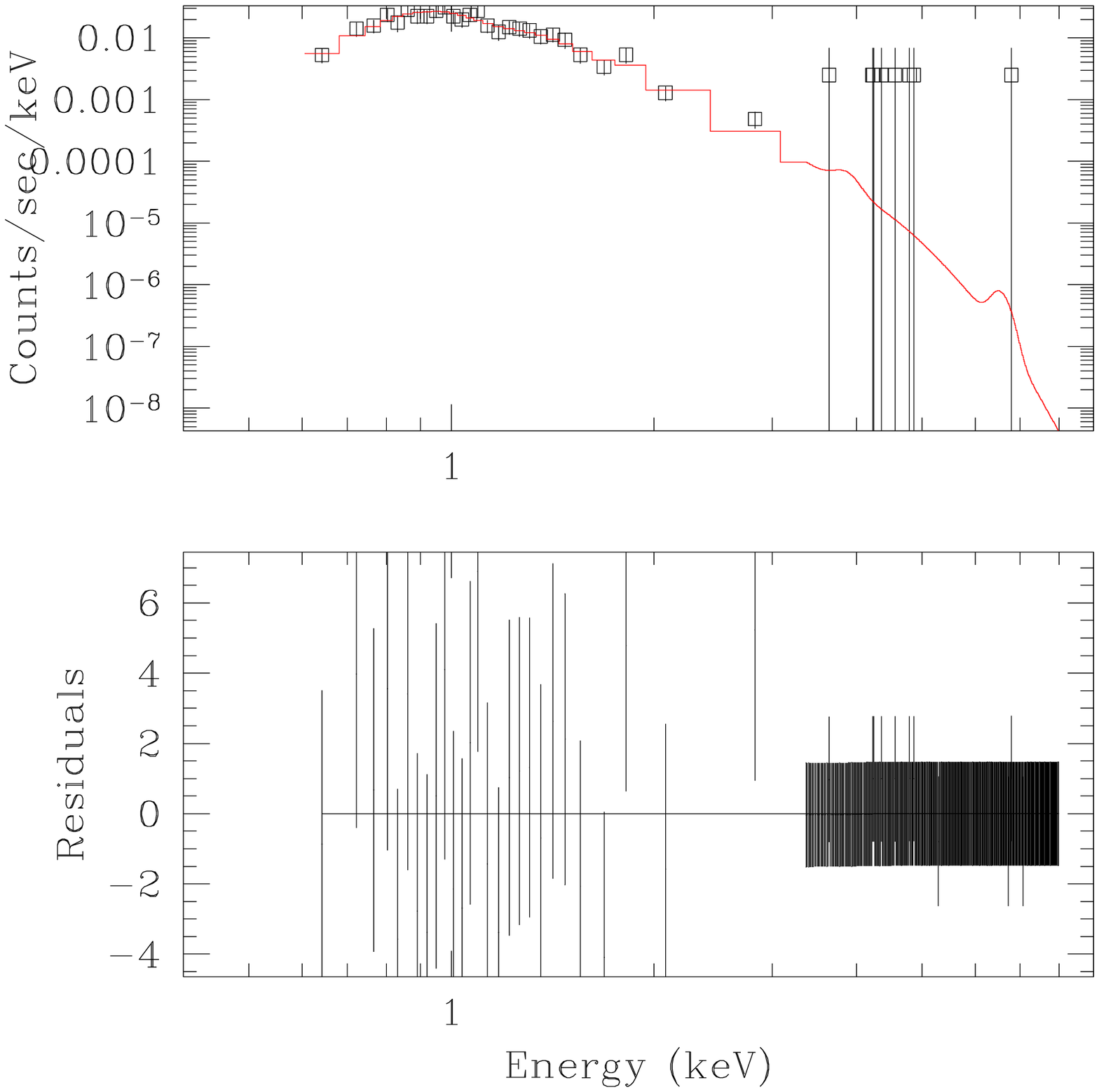}
\caption{Results of {\tt SHERPA} model fit to the {\it Chandra} spectrum
of star 116 from the first Garmire exposure.
\label{fig-star116-g1}}
\end{figure}

\begin{figure}
\plotone{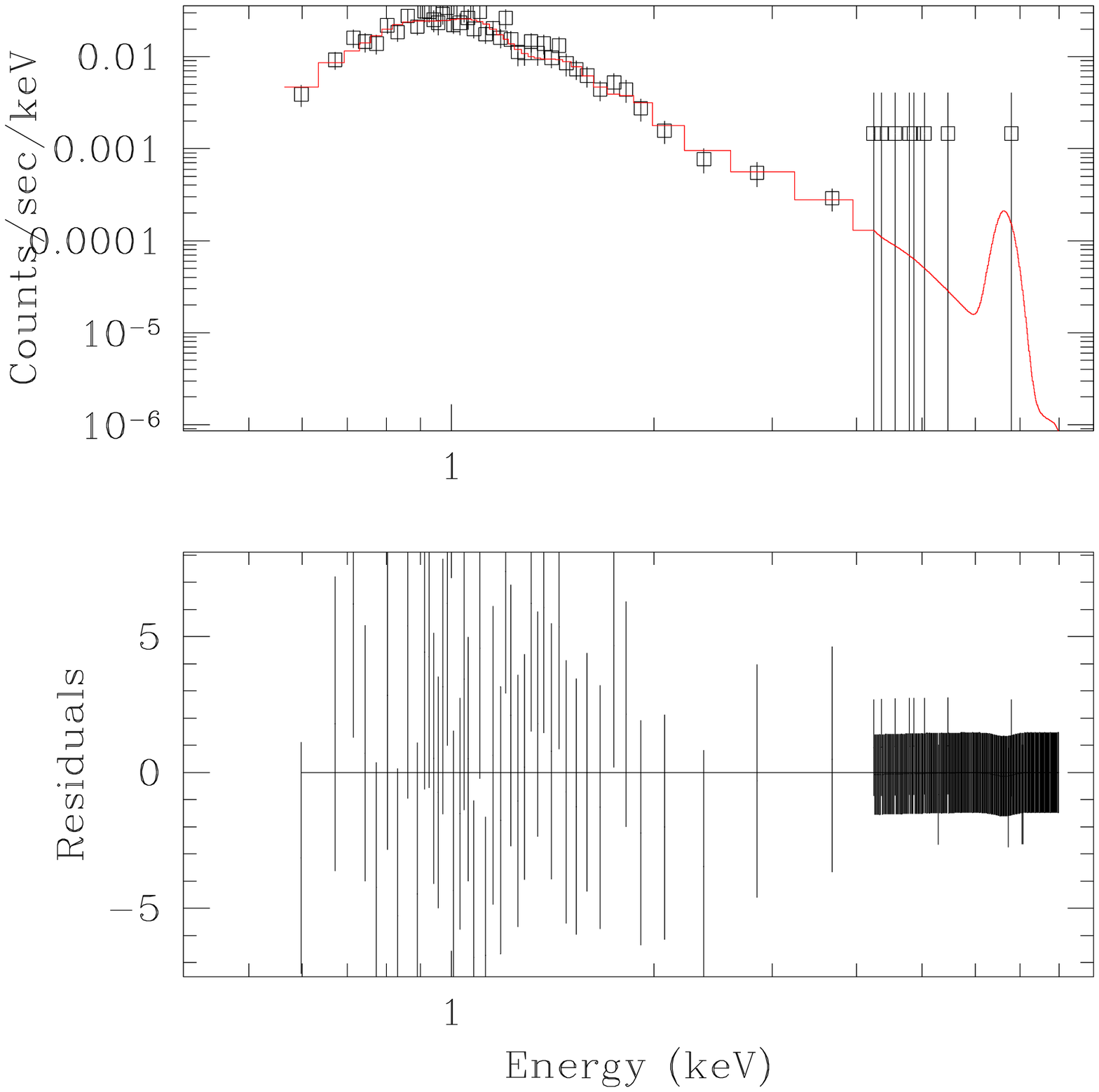}
\caption{Results of {\tt SHERPA} model fit to the {\it Chandra} spectrum
of star 116 from the first Garmire exposure using pre-filtered data.
\label{fig-star116-test}}
\end{figure}

\begin{figure}
\plotone{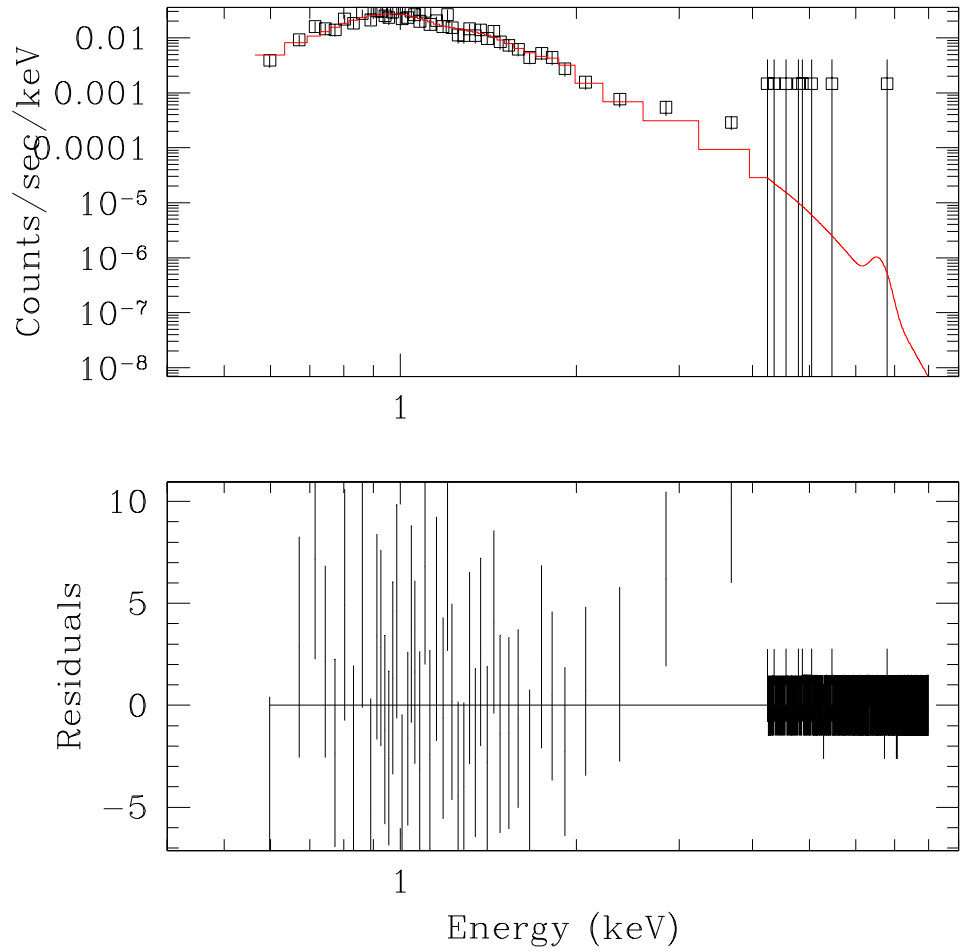}
\caption{Results of {\tt SHERPA} model fit to the {\it Chandra} spectrum
of star 116 from the first Garmire exposure using pre-filtered data and
a single-component thermal plasma model.
\label{fig-star116-test2}}
\end{figure}

Close inspection of our {\tt SHERPA} fits to the two observations of this
source (Fig.\ \ref{fig-star116-g1}) do not indicate any obvious problems.
Perhaps the discrepancy is the result of our flare filtering procedure.
However, the light curve of this source does not include any strong flares
and so was not heavily filtered. In any case, we performed the {\tt SHERPA}
analysis once again but on the pre-filtered data from the first Garmire
exposure. As expected, the resulting 
$L_X$ of $10^{30.5}$ erg/s differs only slightly from the value we report in 
Table \ref{table-lx},
and the model fit again does not present any obvious problems 
(Fig.\ \ref{fig-star116-test}). Recalling that the \citet{feig02} 
analysis typically used single-component fits to
the spectra as compared to our two-component fits, we attempted to
reproduce their value by again running the {\tt SHERPA} analysis on the
pre-filtered data but this time using only one thermal plasma component 
to the model fit. The value of $L_X$ that we derive here ($10^{30.4}$ erg/s)
still does not resolve the discrepancy, and may in fact be a low
measure as the model fit in this case underestimates the flux in the
two highest energy bins that are not upper limits (Fig. \ref{fig-star116-test2}). 

Thus in this example case, and in the other discrepant cases seen in
Fig.\ \ref{fig-lxcomp}, we are simply unable to determine the cause of the
discrepancy. We provide this exercise as a
cautionary lesson about the limits inherent in this type of analysis,
but take comfort in the fact that for the majority of
the sources used in our analysis the agreement between our values and those
derived by \citet{feig02} is in fact very good. 

\section{Results\label{results}}
The X-ray luminosities for each source in Table \ref{table-sample} resulting
from our analysis are given in Table \ref{table-lx},
representing 220 stars with known rotation periods that are included in the
optical database of \citet{hill97}.
In this section we report the results for the 154 sources 
having a quality flag of `1'. 
We remind the reader that our values of $L_X$ are broadband luminosities
over the energy range 0.5 keV to 8 keV, are not corrected for absorption,
and do not include photon events that occur during a flare
(see \S \ref{analysis}).
\citet{feig02} report $L_X$ measurements for an additional 63 stars with rotation
periods detected at lower signal-to-noise; where appropriate we include
these measurements in our analysis and discussion, but in all cases we
maintain a distinction between this larger sample and the subset which we
believe to be of highest quality.

We begin by presenting the basic X-ray properties of these sources, emphasizing
two biases that appear to be inherent to PMS stars having
measurable rotation periods (\S \ref{basic}), namely, a tendency toward
higher X-ray luminosities (\S \ref{lx}) and toward higher levels of X-ray
variability (\S \ref{var}). With these biases in mind, 
we next examine the X-ray data vis-a-vis rotation (\S \ref{rotation})
and accretion (\S \ref{accretion})
for clues into the possible mechanisms for X-ray production in these stars. 

\subsection{Basic X-ray properties of stars with known rotation periods
\label{basic}}
In this section we discuss the basic X-ray properties---luminosity and
variability---of stars with known rotation periods.
By comparing these properties to those of other stars detected in the 
{\it Chandra} observations, we find two biases---astrophysical in origin---in 
the rotation-period sample. 
Stars with measured rotation periods are:
(1) more X-ray luminous both absolutely (i.e.\ $L_X$) and relative
to the stellar bolometric luminosity (i.e.\ $L_X/L_{bol}$);
and (2) more likely to be X-ray variable than are stars in the overall 
PMS population of the ONC. 
These results are highly statistically significant. 
We emphasize that these biases are not due to observational bias (e.g., 
optical magnitude bias) 
in the rotation-period sample, and are therefore likely to
have a physical basis as we discuss in \S \ref{disc-rotation}.
Here we present the evidence for these two biases in turn.

\subsubsection{Bias: X-ray luminosity\label{lx}}
\begin{figure}
\epsscale{0.8}
\plotone{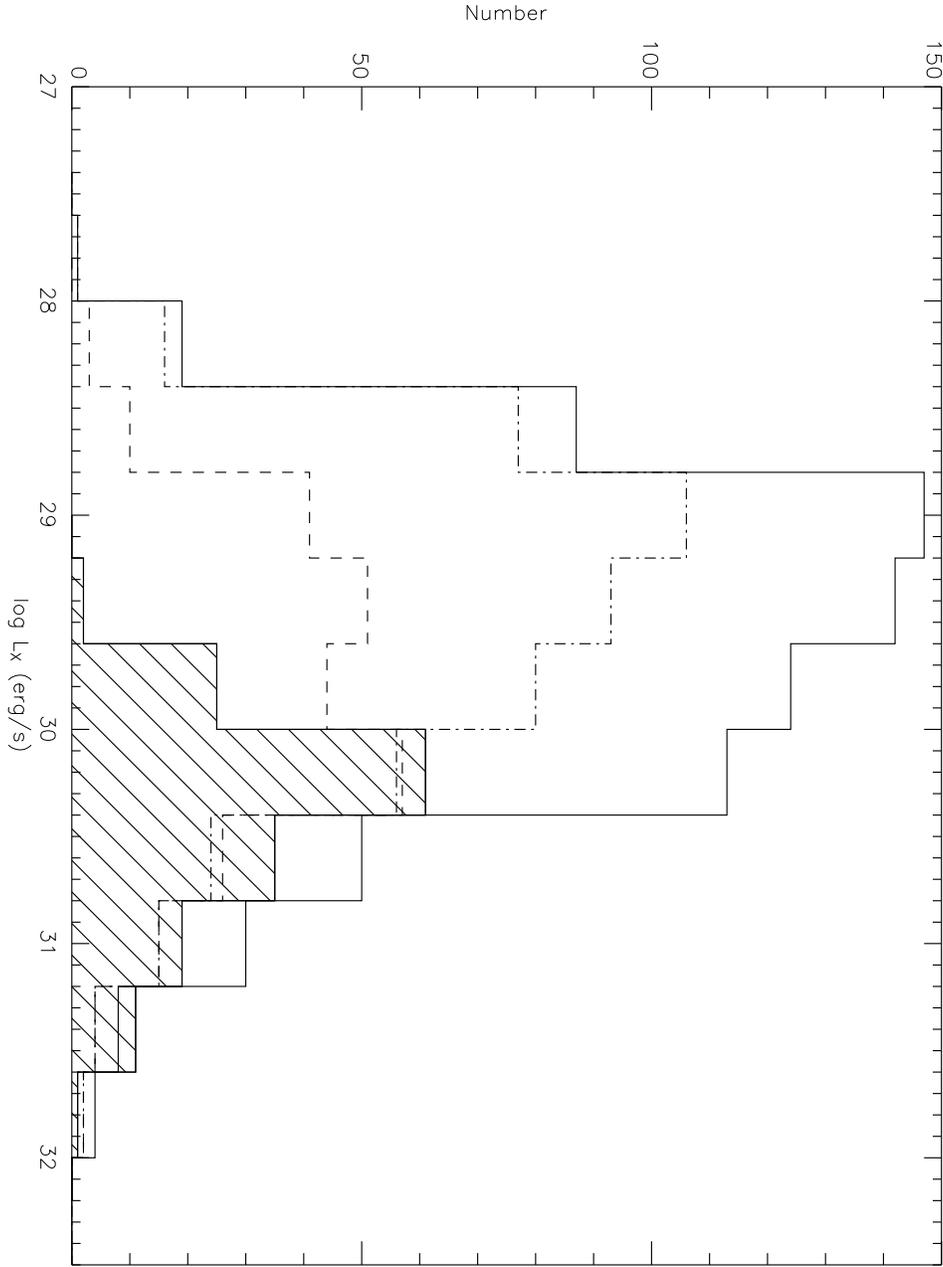}
\caption{Distribution of $\log L_X$ for all ONC stars with optically determined
rotation periods detected by \citet{feig02} (dashed) and the distribution for
those stars with high signal-to-noise detected in this study (hatched).
For comparison, the solid histogram shows the distribution for all ONC stars 
detected by \citet{feig02} having optical magnitudes bright enough ($I \lesssim 17$)
to have been included in the optical rotation-period surveys of the ONC
\citep{stass99,herbst02}.
The distribution for stars lacking rotation period measurements are indicated 
by the dot-dashed histogram.
Stars with optically determined rotation periods are systematically 
biased to higher $L_X$ as compared to the underlying population.
\label{fig-lxbias}}
\end{figure}

We find that ONC stars with known rotation periods are significantly 
biased to high X-ray luminosities. In Fig.\ \ref{fig-lxbias} we plot both
the distribution of $\log L_X$ for our study sample (Table \ref{table-lx};
hatched histogram) as well as the larger sample of stars with
rotation periods detected by \citet{feig02} (dashed histogram).
For comparison, the solid histogram shows
the distribution of $\log L_X$ for all stars reported by \citet{feig02}
included in the optical survey of \citet{hill97}. To demonstrate that the
bias to high $L_X$ among stars with rotation periods is not due to optical
bias in the rotation-period studies, we include here only those stars detected 
by \citet{feig02} having optical 
magnitudes bright enough ($I \lesssim 17$) to have been included in
the optical samples studied for rotation periods \citep{stass99,herbst02}.
We further restrict this comparison sample to only stars with masses
$M < 3\; {\rm M}_\odot$, as this represents the range of stellar masses
among stars with rotation period measurements.

While the stars with known rotation periods (dashed histogram) exhibit a 
range of $L_X$, this range is $\sim 0.5$ dex smaller than that spanned
by the underlying ONC population (solid histogram). Moreover, the $L_X$ distribution 
of these stars is skewed with respect to the overall distribution, 
such that stars with rotation periods exhibit higher average $L_X$. 
To show this more clearly, the distribution of $L_X$ for stars {\it without}
rotation periods (i.e.\ the difference between the solid and dashed histograms)
is shown also (dot-dashed histogram).

A two-sided K-S test indicates that the probability of the $L_X$ distributions 
for stars with and without rotation periods (dashed and dot-dashed histograms)
being drawn from the same parent population is $7 \times 10^{-7}$.
In addition, a Student's $t$ test gives a probability of only $2 \times 10^{-10}$
that the means of these two distributions ($\log L_X = 29.75$ erg/s
for stars with rotation periods and $\log L_X = 29.39$ erg/s for stars
without) are the same.

A similar result is obtained when we consider $L_X / L_{bol}$ instead of $L_X$
(Fig.\ \ref{fig-lxlbolbias}). Here, a two-sided K-S test gives a probability
of $2 \times 10^{-8}$
that the $L_X / L_{bol}$ distributions for stars with and without rotation
periods are drawn from the same parent population. 
And a Student's $t$ test gives a probability of $6 \times 10^{-10}$ that the
means of these two distributions ($\log L_X / L_{bol} = -3.67$ for stars
with rotation periods and $\log L_X / L_{bol} = -4.09$ for stars without)
are the same.

\subsubsection{Bias: X-ray variability\label{var}}
A similar bias manifests itself with respect to X-ray variability
of the sources. 
The subset of stars in our sample whose X-ray light 
curves are variable (`Flare', `PosFl', or `LTVar' in Table \ref{table-lx}) 
comprise 82\% $\pm$ 3\% of our study sample (uncertainties determined
from the binomial distribution). 
Similarly, 70\% $\pm$ 3\% of stars with rotation periods in the larger sample of 
\citet{feig02} show variability.
In comparison, a smaller fraction, 57\% $\pm$ 2\%, of 
ONC stars in the \citet{feig02} study that lack rotation periods show such
variability. 

For the entire sample of stars with rotation periods, this difference 
in X-ray variability is statistically significant. A $\chi^2$ test gives
a probability of 0.001 that stars with and without rotation periods have
equal occurrences of variability.
For our high-quality sample, where the signal-to-noise is higher and
variability in the light curves is therefore better determined, a
$\chi^2$ test gives a probability of $2\times 10^{-9}$ that the occurrence of
variability is the same as that found among stars without rotation periods.

There thus appears to be significant evidence for an enhancement of X-ray 
variability among stars in the ONC with rotation periods, particularly when we 
restrict our analysis to those stars with the highest quality X-ray 
light curves. 

\subsection{Rotation\label{rotation}}
\begin{figure}
\plotone{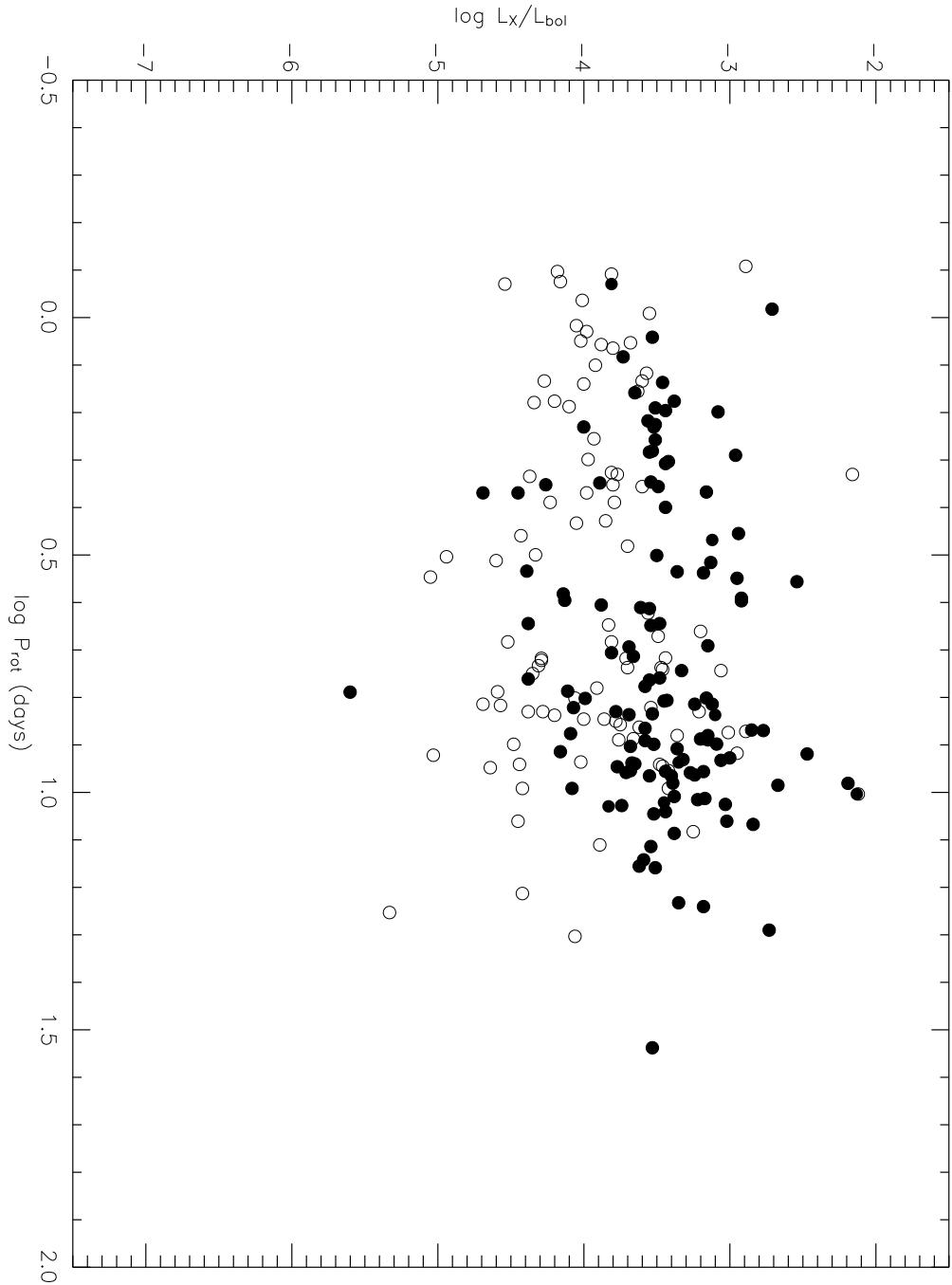}
\caption{The X-ray/rotation relationship for PMS stars in the ONC with
known rotation periods. Filled circles represent stars detected in this
study, open circles represent additional stars from the study of \citet{feig02}.
\label{fig-lxvsrot}}
\end{figure}

X-ray emission on the main sequence among stars with $M \lesssim 3$ M$_\odot$ is 
believed to be driven by stellar
rotation, and this results in a clear, observable correlation between stellar
rotation and X-ray luminosity.
The relationship between X-ray luminosity and stellar rotation period
for our study sample is shown in Fig.\ \ref{fig-lxvsrot}, where we
plot $\log L_X / L_{bol}$ vs.\ $\log P_{rot}$. 
For ease of comparison, the
vertical scale is set to the full range of $\log L_X / L_{bol}$ observed 
on the main sequence.

As noted above and in the previous studies of \citet{flac-acc} and 
\citet{feig03}, these stars show a mean $\log L_X / L_{bol}$ near the main 
sequence saturation value of $-3$, though somewhat lower 
(mean $\log L_X / L_{bol} = -3.67$ for all stars with rotation periods).
Taken at face value, these data present no clear evidence for an 
X-ray/rotation relationship of the sort seen on the main sequence. 

At a more detailed level, these data provide possible evidence for these
stars being in the super-saturated regime of the rotation/X-ray 
relationship. In addition to having a mean $L_X/L_{bol}$ below the
saturation value, the data in Fig.\ \ref{fig-lxvsrot} also
show a weak, but statistically significant, trend of increasing 
$L_X / L_{bol}$ with increasing rotation period, as might be expected 
for stars in the super-saturated regime.
Among all stars with rotation periods,
a Spearman's $\rho$ rank-correlation test gives a probability
of $9 \times 10^{-4}$ that $P_{rot}$ is uncorrelated with $L_X/L_{bol}$.
The same trend is present among the smaller set of stars detected 
in this study, though only at 95\% significance.

\begin{figure}
\plotone{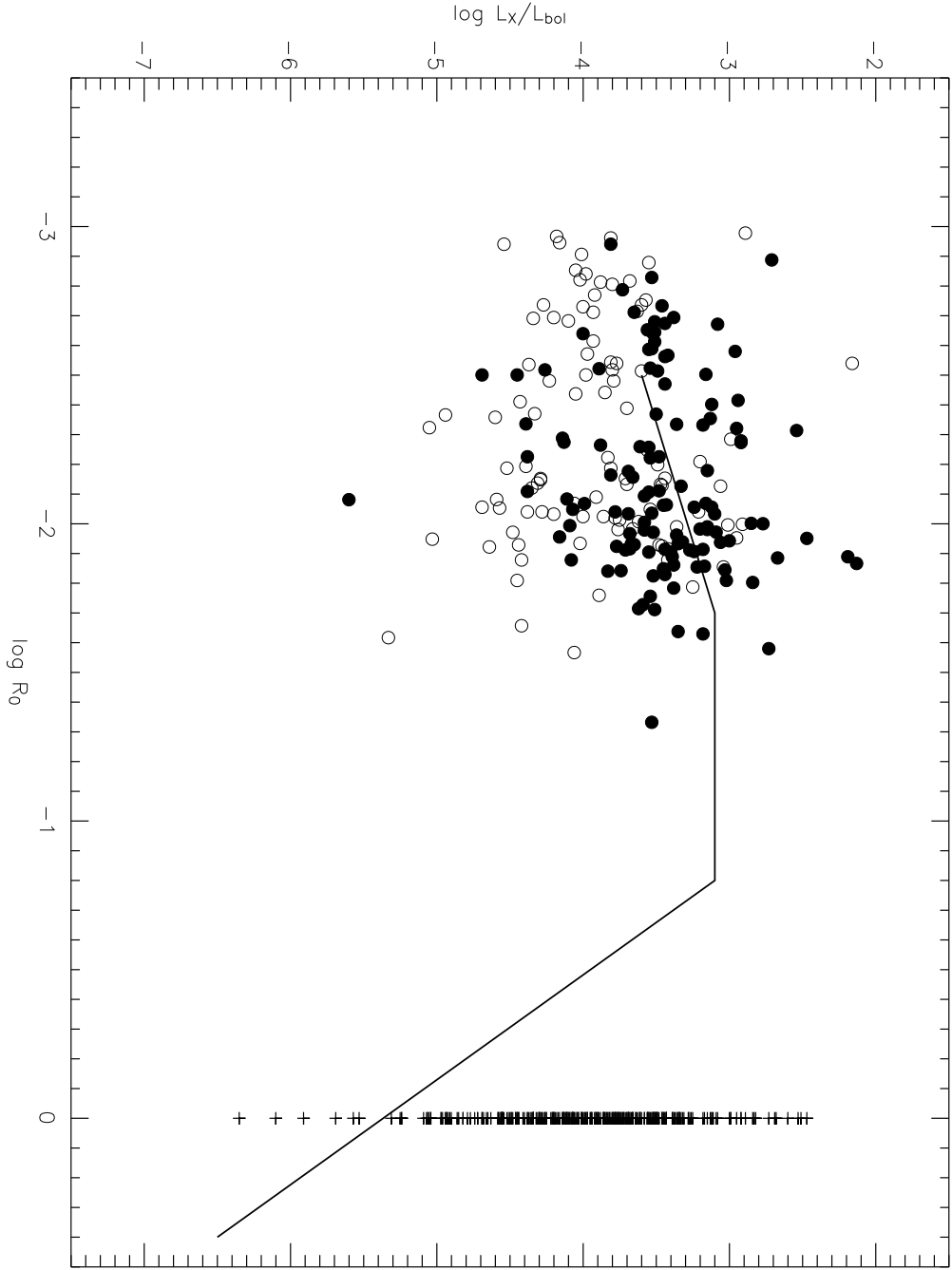}
\caption{The X-ray/rotation relationship for PMS stars in the ONC with respect
to Rossby number instead of $P_{rot}$. Point symbols are as in Fig.\ \ref{fig-lxvsrot}.
The main sequence relationship is
indicated by the solid line for comparison. Also shown 
(crosses) is the remainder of the sample included in the study of \citet{feig02}
with $M < 3\; {\rm M}_\odot$ (plotted arbitrarily at $\log R_0 = 0$).
\label{fig-lxvsrossby}}
\end{figure}

To effect a better comparison with super-saturation on the main sequence, 
we transform the abscissa from
$P_{rot}$ to Rossby number, $R_0$, defined as the ratio between $P_{rot}$
and the convective turnover timescale, 
$\tau_c$\footnote{The convective turnover timescale, $\tau_c$, is typically 
determined from stellar interiors models for stars of the appropriate
mass and age. As discussed by \citet{flaccomio-thesis}, at the young age
of the ONC the value of $\tau_c$ is roughly constant for these fully
convective low-mass stars. We thus convert $P_{rot}$ to $R_0$ by 
scaling the former by a constant value of $\tau_c = 800$ days 
\citep{ventura}.}, which is typically used to show the X-ray/rotation 
relationship on the main sequence.
This is shown in Fig.\ \ref{fig-lxvsrossby}, where
the solid line represents the main-sequence relationship as
determined by \citet{pizzolato}, and where
the stars in our sample now appear explicitly in the
super-saturated regime. 

\begin{figure}
\epsscale{0.85}
\plotone{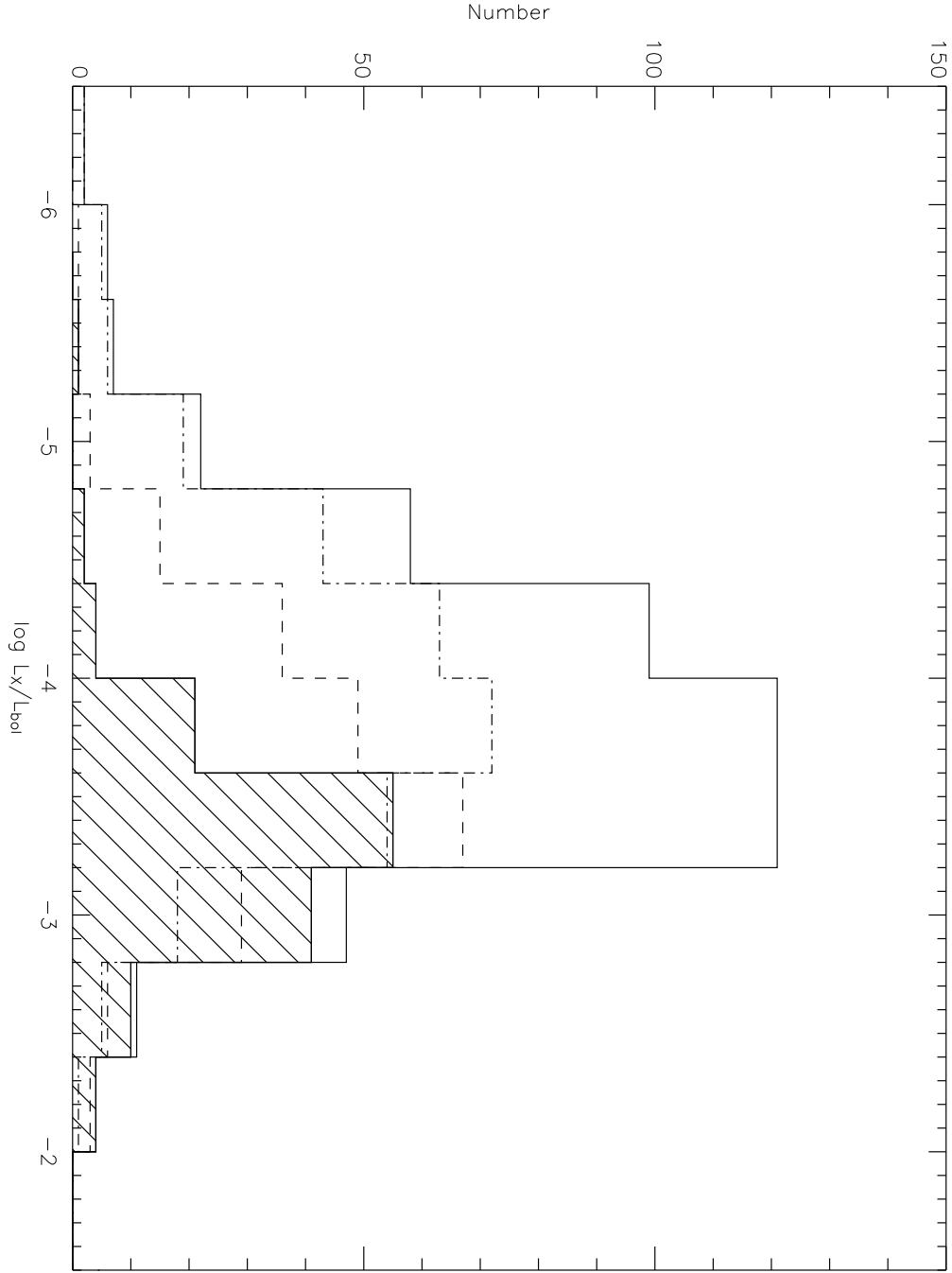}
\caption{Same as Fig.\ \ref{fig-lxbias}, except showing $L_X/L_{bol}$ instead
of $L_X$. The bias for stars with rotation periods toward higher $L_X$ 
is evident in $L_X/L_{bol}$ also.
\label{fig-lxlbolbias}}
\end{figure}

Fig.\ \ref{fig-lxvsrossby} also shows the
$\log L_X / L_{bol}$ for the remainder of the ONC sample from \citet{feig02}
with $M < 3\; {\rm M}_\odot$ (crosses plotted arbitrarily at 
$\log R_0 = 0$; these are the same stars as in the dot-dashed histogram in 
Fig.\ \ref{fig-lxlbolbias}). 
As we have seen (\S \ref{lx}, Figs.\ \ref{fig-lxbias} and \ref{fig-lxlbolbias}), 
these stars are on average less X-ray
luminous than are stars with known rotation periods.
Might there also be differences on average in their rotational properties? 

\begin{figure}
\plotone{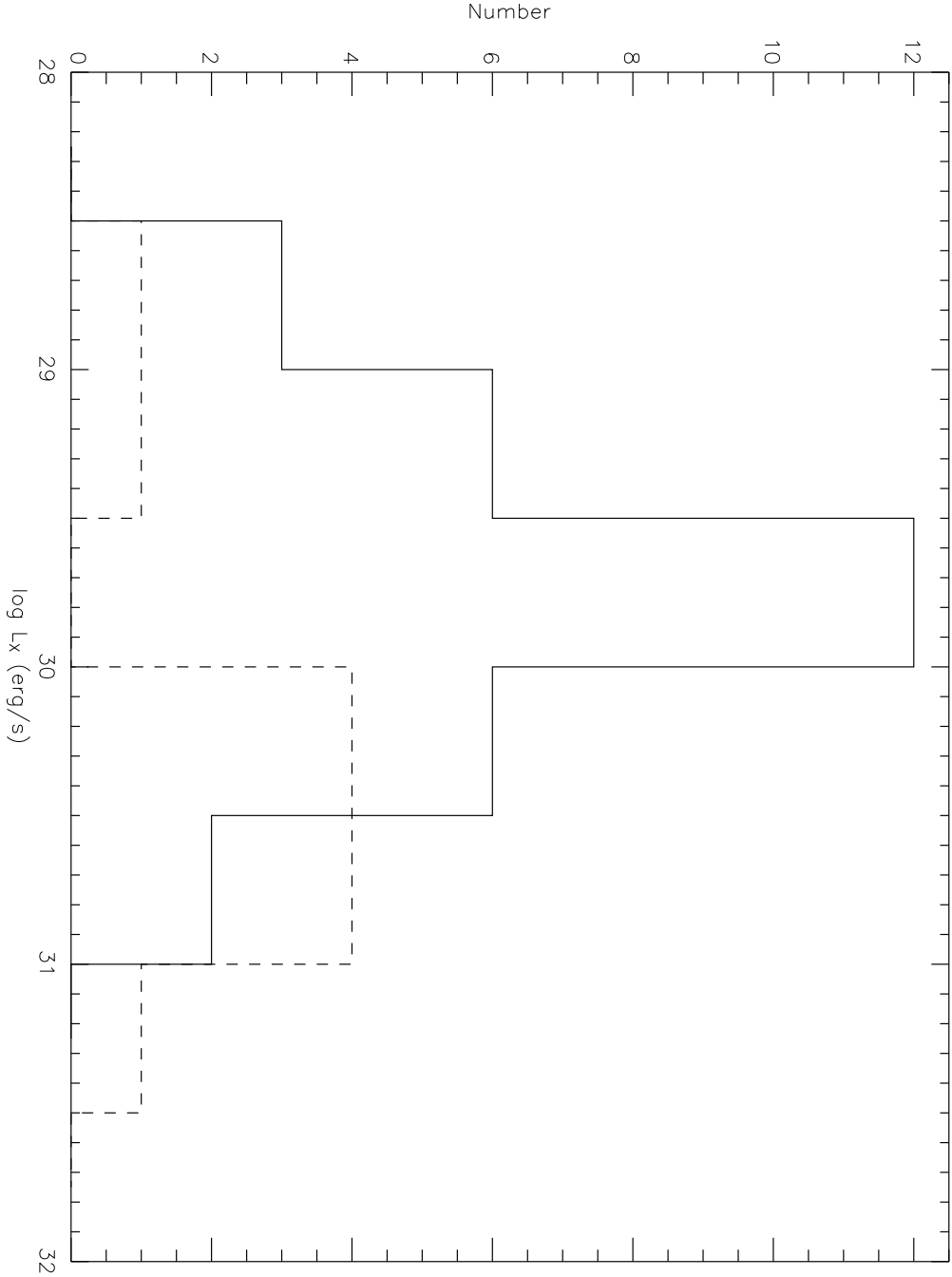}
\caption{Distribution of $\log L_X$ for stars lacking $P_{\rm rot}$ but
with $v \sin i$ measurements from \citet{rhode}. The solid histogram
represents slow rotators, defined as stars with $v \sin i$ upper limits,
whereas the dashed histogram represents rapid rotators, defined as stars
with broadened spectral lines. \citet{rhode} report an instrumental 
resolution of $\approx 14$ km/s.
\label{fig-vsini}}
\end{figure}

For 40 of these stars lacking optical rotation periods, $v \sin i$ 
measurements are available from the study of \citet{rhode}, allowing us to 
infer their (projected) rotational characteristics. In Fig.\ \ref{fig-vsini} 
we show the $L_X$ distribution for these stars segregated into two groups, 
fast (11 stars) and slow rotators (29 stars), defined on the basis of whether 
\citet{rhode}
report a $v \sin i$ measurement or a $v \sin i$ upper limit (i.e.\ whether
the spectral lines are broadened beyond the instrumental resolution or not).
The slow rotators indeed appear to be skewed to lower $L_X$, and both a two-sided 
K-S test and a Student's $t$ test confirm this at the 99\% confidence level. 
The difference between slow and rapid rotators is not statistically
significant when we consider $L_X/L_{bol}$ instead of $L_X$.

A similar test is possible among stars with $v\sin i$ measurements that
{\it do} have rotation periods (58 fast and 62 slow rotators).
The $L_X$ distributions of these two groups are statistically indistinguishable.
Apparently, the difference in $L_X$ between fast and slow rotators is only
present among stars lacking optical rotation periods.

Thus, while there is not a one-to-one correlation between $L_X$ and
$v \sin i$ for stars without optical rotation periods, there is a
marginally significant tendency for the X-ray faint stars in this
group to also have slower rotation speeds. This is in
the opposite sense to what we find above for stars that do have optical 
rotation periods, in which the X-ray luminosity {\it increases} with
slower rotation similar to super-saturated stars on the main sequence
(cf.\ Fig.\ 9 in \citet{pizzolato}), albeit with a large scatter.

\subsection{Accretion\label{accretion}}
Accretion is another mechanism possibly related to X-ray production in
PMS stars, and indeed accretion appears to manifest itself strongly 
in the X-ray properties of the stars in our study. 
We use the strength of
emission in the \ion{Ca}{2} line as measured by \citet{hill98} to
determine which stars are actively accreting: Following \citet{flac-acc}
we take stars with \ion{Ca}{2} equivalent widths (EW) of $< -1$ \AA\ 
(i.e.\ in emission) to be those actively accreting, while those with
EW $> 1$ \AA\ (i.e.\ in absorption) to be non-accreting. 

\ion{Ca}{2} EW measurements are available for 117 stars in our sample
and for 199 stars among all stars with rotation periods. 
In light of the biases inherent to the rotation period sample noted
in \S \ref{basic}, where appropriate we also explore accretion signatures
in the full sample of ONC stars from the study of \citet{feig02}.

We find that stars with active accretion signatures in \ion{Ca}{2},
while no more likely to show X-ray flares than non-accreting stars, are
systematically less X-ray luminous and exhibit systematically harder X-ray spectra. 
We discuss in turn the relationship between accretion and X-ray flaring,
X-ray luminosity, and X-ray hardness.

\subsubsection{Accretion and X-ray flaring\label{acc-flaring}}
We begin by noting that spectroscopic signatures of active accretion
are relatively rare among the stars in our sample. Among the 117 stars
from this study that have \ion{Ca}{2} measurements, only 10 stars show
\ion{Ca}{2} clearly in emission (i.e.\ EW $< -1$ \AA), whereas 66 stars show
\ion{Ca}{2} clearly in absorption (i.e.\ EW $> 1$ \AA).  Among those few
stars that do show evidence for active accretion, all 10 of them exhibit
X-ray flaring in the {\it Chandra} data (`Flare' or `PosFl' in Table \ref{table-lx}).
Among the non-accreting stars, 70\% (46/66 stars) show such evidence 
for X-ray flaring.  Because of the small number of accreting sources in
this sample, this difference is not statistically significant.

Similarly, among the larger sample of all stars with rotation periods only 
28/199 stars show
\ion{Ca}{2} clearly in emission, whereas 77 stars show \ion{Ca}{2} in absorption.
Among the 28 accreting stars, 15 (54\%) show evidence for X-ray flaring,
while among the non-accreting stars 47 stars (61\%) do. This small difference 
is not statistically significant.

Considering the entire ONC sample included in the study of \citet{feig02},
there are 254 stars for which \citet{hill98} report a \ion{Ca}{2} EW
of either $< -1$ \AA\ (126 stars) or $> 1$ \AA\ (128 stars). In this
larger sample, 41\% of the accreting stars show X-ray flaring, and 48\%
of the non-accreting stars do, again indicating no relationship between
accretion and X-ray flaring.

We thus find that while stars with optical rotation periods are predominantly
non-accreting (see also \citet{stass99,herbst02}), X-ray flaring is 
nonetheless ubiquitous among them
(\S \ref{var}), and the presence of active accretion does
not significantly enhance this X-ray flaring.

\subsubsection{Accretion and X-ray luminosity\label{acc-lum}}
Among the stars with measured rotation periods, we find a hint that
actively accreting stars have lower X-ray luminosities than their
non-accreting counterparts. As above, there are only 28 stars with
rotation periods that show clear signs of active accretion and 77
stars that clearly do not. Comparing the $L_X$ distributions of these
two subsets, a Student's $t$ test reveals different means---with 
accretors being less luminous---at 98\% confidence.

\begin{figure}
\epsscale{0.75}
\plotone{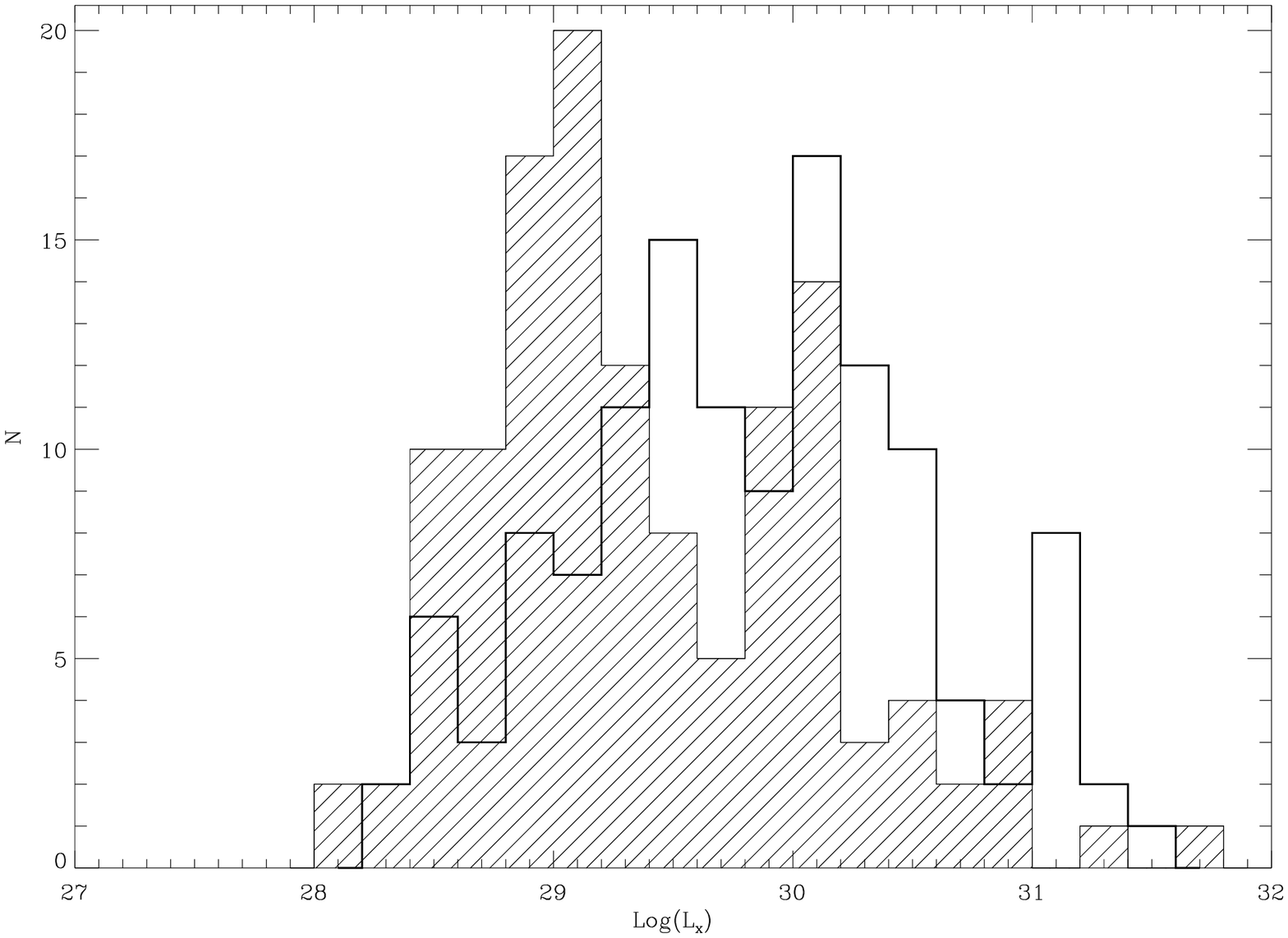}
\plotone{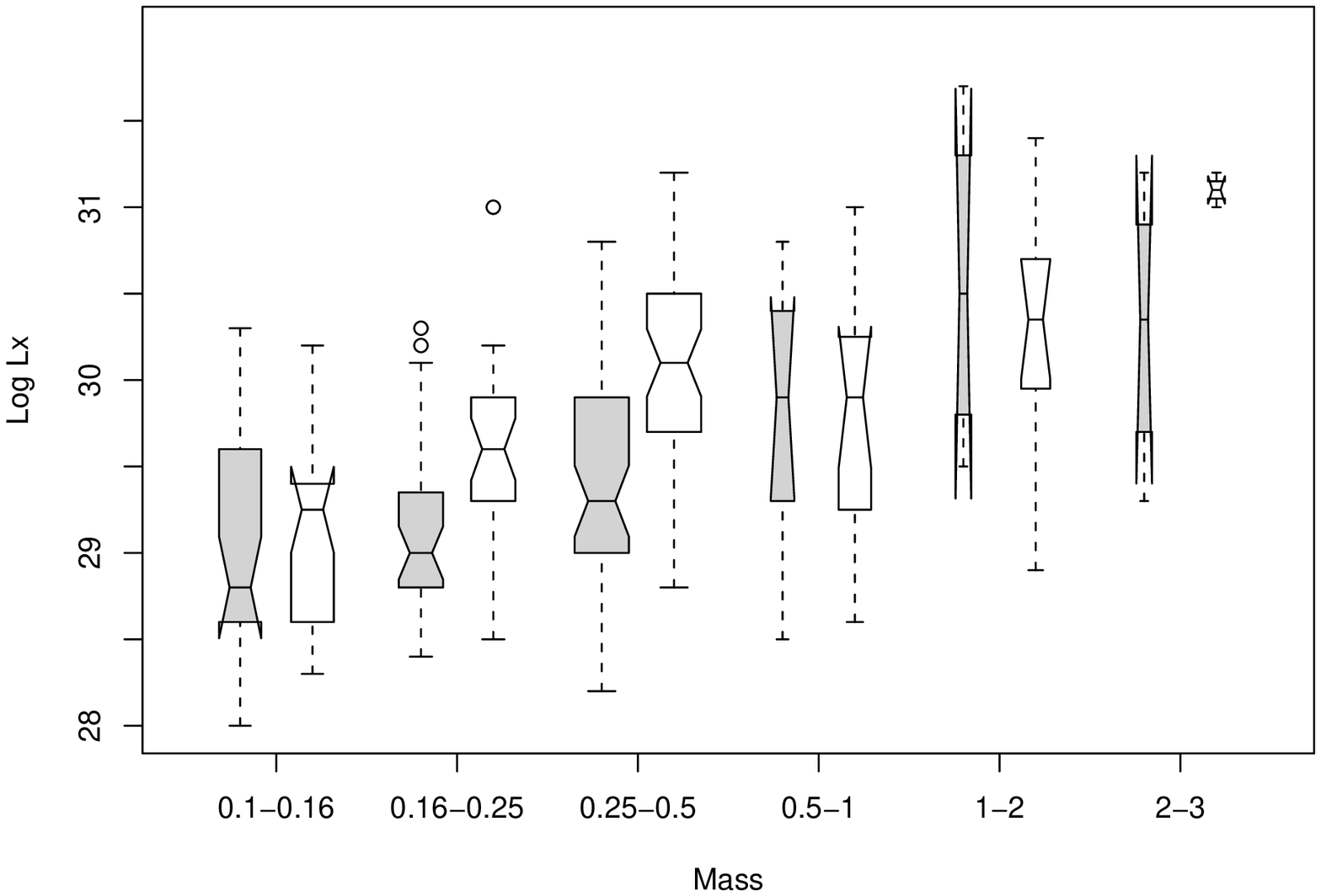}
\caption{\label{lx-acc} (Top) Distribution of X-ray luminosities for the
\citet{feig02} data. The hatched (clear) histogram is for stars with  
EW(\ion{Ca}{2}) $<-1$ \AA\ (EW(\ion{Ca}{2}) $>1$ \AA). (Bottom) Box plots for X-ray 
luminosities, binned as a function of mass. The gray (clear) boxes correspond to 
accretors (non-accretors). The width of each box is proportional to the 
square root of observations in each bin. The scale of the abscissa is 
arbitrary. For the accretors, there are 29,32,48,9,4, and 4 stars in 
each increasing mass bin. For the non-accretors there are 26,28,42,15,12, 
and 4 stars.}
\end{figure}

However, within the full ONC sample 
we find that this difference in
$L_X$ between accretors and non-accretors is highly statistically significant. 
Of the 529 stars from \citet{hill98} in the {\it ACIS} field, \citet{feig03} 
detect 525 stars with $M < 3\; {\rm M}_\odot$. Of these,
256 have EW(\ion{Ca}{2}) $< -1$ \AA\ (126 detected in X-rays, 0 undetected) or
EW(\ion{Ca}{2}) $> 1$ \AA\ (128 detected in X-rays, 2 undetected). Here
we ignore the two undetected stars. As Fig.\ \ref{lx-acc}a shows, the $L_X$
distributions of accretors and non-accretors are clearly different; a
two-sided K-S test reveals that the probability that the two are drawn
from the same parent distribution is $3\times 10^{-5}$.

As demonstrated by \citet{feig02} and \citet{flac-basic}, $L_X$ correlates
strongly with stellar mass. Thus, the differences in $L_X$ among accretors
and non-accretors might be the result of a correlation between accretion
and stellar mass. Fig.\ \ref{lx-acc}b shows the $L_X$ distributions for
accretors and non-accretors as a function of mass (stellar masses taken
from \citet{feig02}). The center of each box
markes the position of the median $L_X$ in that mass bin. If the indented
regions around the medians (``notches") of two boxes do not overlap, the
medians are different with $>$ 95\% confidence (see \citet{feig03} for an
explanation of box plots). We see that for stars below $\sim 0.5$ M$_\odot$,
those with spectroscopic accretion indicators have significantly lower
$L_X$ than stars that do not have spectroscopic accretion indicators.
The number of objects in the higher mass bins, particularly those 
showing active accretion in \ion{Ca}{2}, is sufficiently small that the
uncertainties on the boxes in Fig.\ \ref{lx-acc}b are large and any 
differences between accretors and non-accretors may be difficult to detect.

\subsubsection{Accretion and X-ray hardness\label{acc-hardness}}
\begin{figure}
\plotone{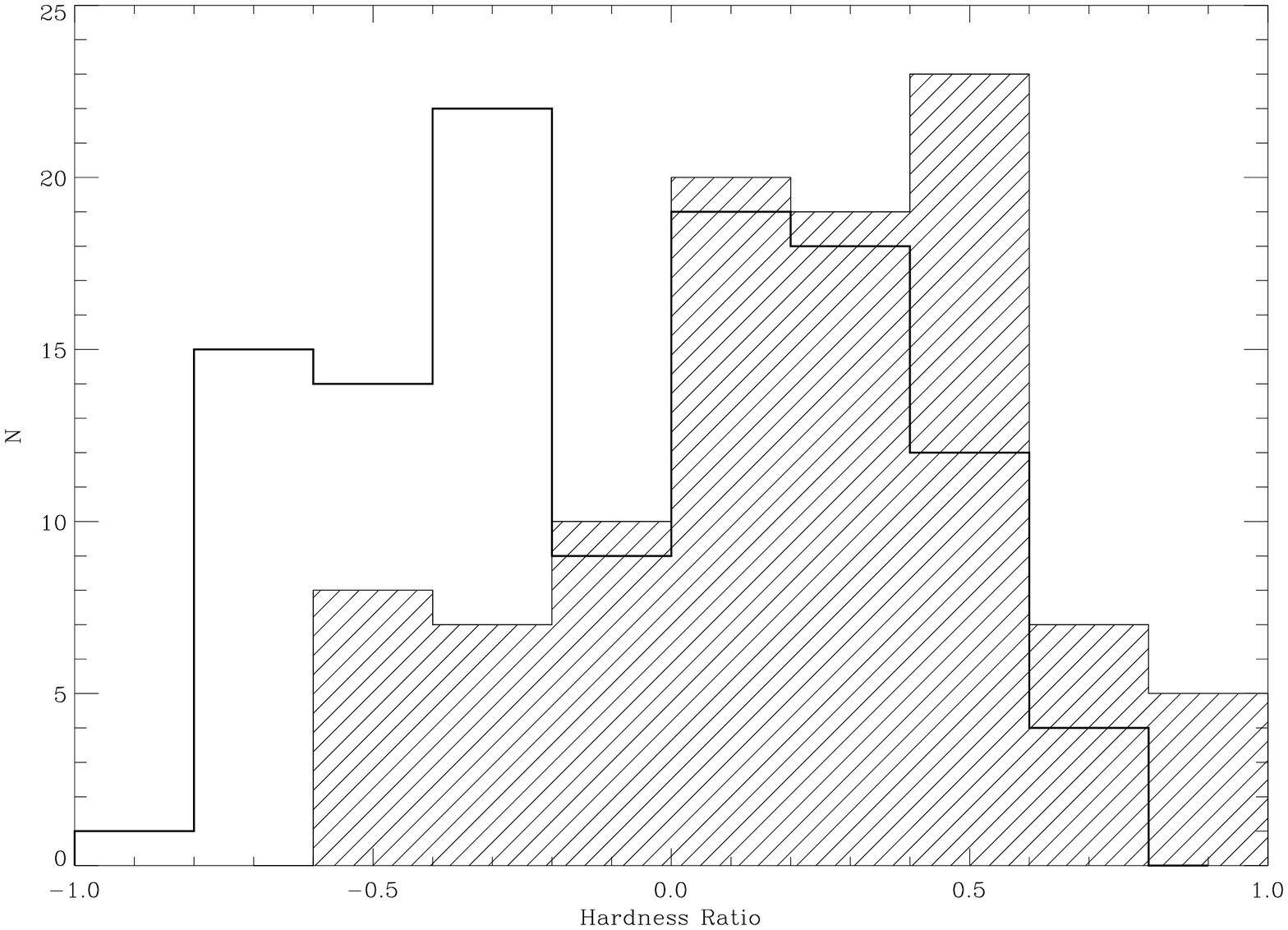}
\plotone{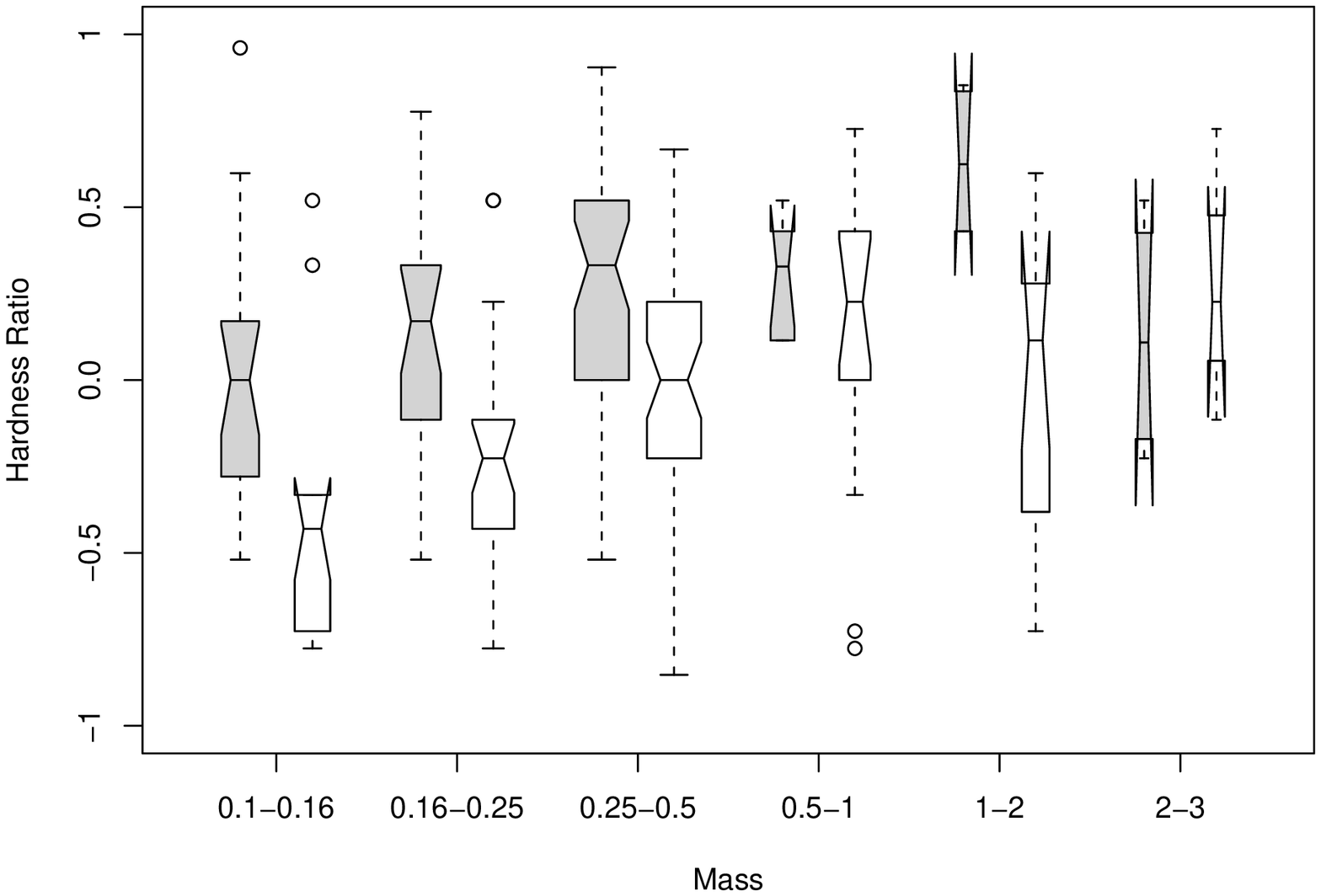}
\caption{\label{hr-acc} (Top) Distribution of HR for \citet{feig02} data. 
The hatched (clear) histogram is for stars with  EW(\ion{Ca}{2}) $< -1$ \AA\
(EW(\ion{Ca}{2}) $>1$ \AA). 
(Bottom) Box plots for HR as a function of mass. As before, gray (clear) 
boxes correspond to accretors (non-accretors).}
\end{figure}

In addition to X-ray luminosity, the {\it Chandra/ACIS} data allow us to
compare accretors and non-accretors in terms of X-ray spectral properties.
Fig.\ \ref{hr-acc}a compares the histograms of hardness ratios
[HR $=(L_h - L_s)/(L_h + L_s)$] for accretors and non-accretors, where 
$L_s$ is the X-ray luminosity
from 0.5 to 2 keV, and $L_h$ is the X-ray luminosity from 2 to 8 keV.
As above, we include in our analysis all ONC stars from the study of
\citet{feig02} with $M < 3$ M$_\odot$.

We find that accretors exhibit systematically harder X-ray spectra than
non-accretors, and the likelihood of both samples being drawn from the
same parent distribution is $10^{-5}$.
Fig.\ \ref{hr-acc}b shows the mass dependence of the HR. Similar to 
Fig.\ \ref{lx-acc}b, a difference between accretors and non-accretors is
clear for stars with masses below $\sim 0.5$ M$_\odot$. 

\section{Discussion\label{discussion}}
From our analysis of all archival {\it Chandra/ACIS} observations of a large
sample of PMS stars in the ONC, we have identified important biases in the
basic X-ray characteristics (luminosity and variability) of stars with 
optically determined rotation
periods as compared to the overall population of PMS stars detected by
{\it Chandra}. In addition, we have explored possible relationships
between the X-rays observed from these stars
and the two physical mechanisms most likely responsible for their
production: rotation and accretion.

In this section we explore in greater depth the implications of the
findings presented in \S \ref{results} toward
the goal of further elucidating the origin of X-rays in PMS stars. 
We structure this discussion again around the two central physical
mechanisms of rotation and accretion. We will argue that the data
hint at the presence of an underlying rotation/X-ray relationship 
qualitatively similar to that observed on the main sequence, and we will
show that the observed differences in X-ray characteristics between 
accretors and non-accretors
are in fact consistent with a picture in which all stars have intrinsically
similar X-ray emission properties. We therefore posit that rotation and
not accretion is primarily responsible for the production of X-rays in
PMS stars at $\sim 1$ Myr.

\subsection{Rotation\label{disc-rotation}}
In seeking to find a rotation/X-ray relationship among PMS stars analogous 
to that observed on the main sequence, it is logical to focus on the 
X-ray properties of PMS stars with known rotation periods.
Unfortunately, the full rotation/X-ray relationship, if it exists among
PMS stars in the ONC, might not be discernable from those stars with
optically determined rotation periods alone.
As we have seen, these stars are significantly biased to higher values of
$L_X$ (and $L_X/L_{bol}$) than are stars without rotation periods. 
These stars may therefore only allow us to probe the 
super-saturated regime of any underlying rotation/X-ray relationship.

Why are PMS stars with optically determined rotation periods biased in
their basic X-ray characteristics? It appears that this bias results 
from the fact that rotation periods can only be 
measured among stars with spots that are sufficiently large and long-lived to
produce stable periodic signals in the optical. 

\begin{figure}
\epsscale{0.9}
\plotone{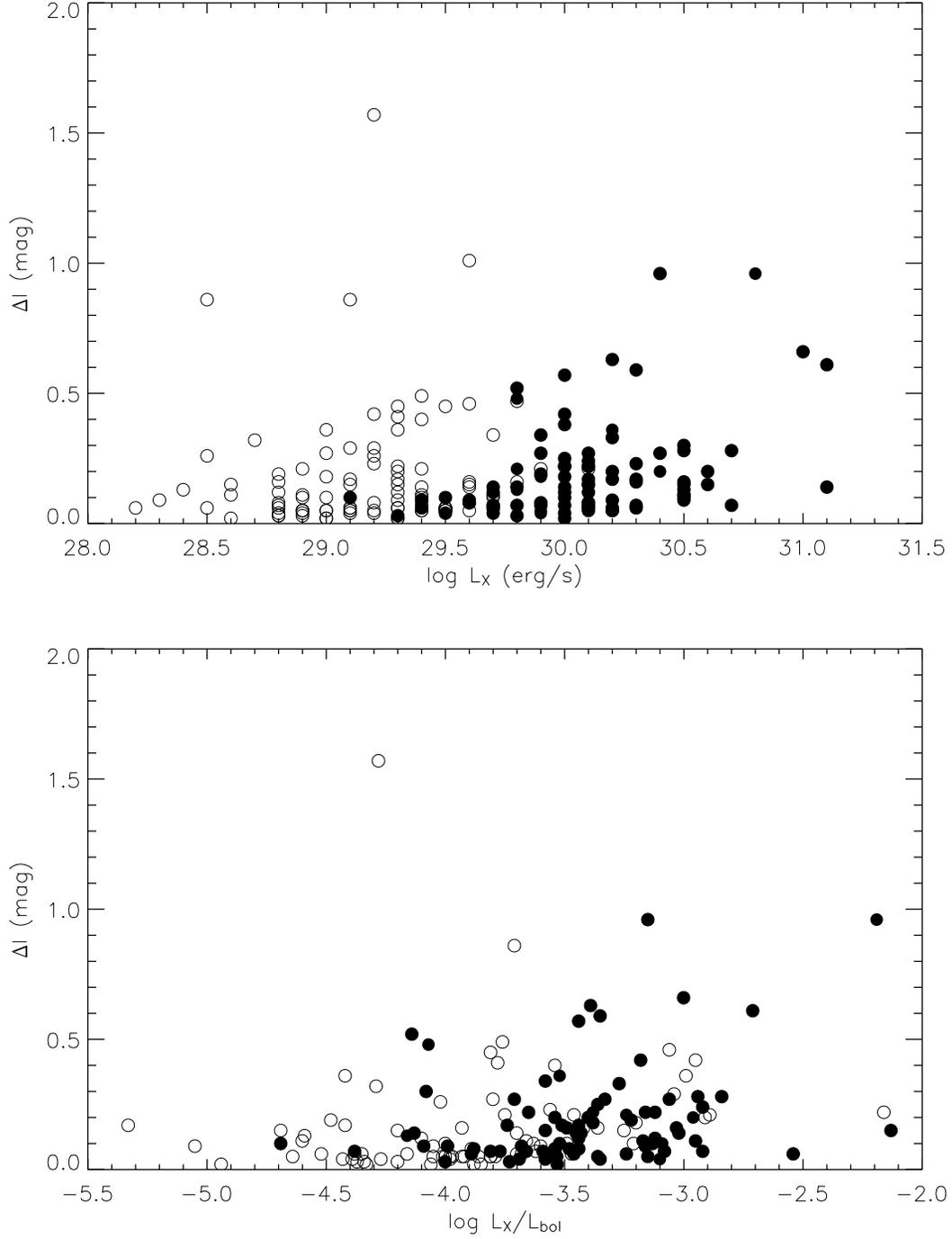}
\caption{Amplitude of photometric variability in the $I$-band is plotted
vs.\ $L_X$ (top) and $L_X/L_{bol}$ (bottom) for stars with optically 
determined rotation periods.
Symbols are as in Fig.\ \ref{fig-lxvsrot}.
\label{lx-spots}}
\end{figure}

To show this,
in Fig.\ \ref{lx-spots}a we plot the amplitude of optical variability, $\Delta I$,
as reported by \citet{herbst02} for PMS stars in the ONC with rotation periods, 
against these stars' X-ray luminosities as determined in this study and in
the study of \citet{feig02}. The two quantities are highly correlated. 
Whether we consider all stars with rotation periods, or only those detected
in this study (filled circles in Fig.\ \ref{lx-spots}), 
a Spearman's rank-correlation analysis yields a probability of 
$\sim 10^{-4}$ that $\Delta I$ and $L_X$ are uncorrelated. 
The same result is obtained when we consider $L_X/L_{bol}$ instead of $L_X$
(Fig.\ \ref{lx-spots}b). In this case, we find a correlation at marginal
confidence (99\%) when we consider only the stars from this study, 
but a probability of $1\times 10^{-6}$ that the two quantities
are uncorrelated when we include all stars with rotation periods.

The implication is that
we do not observe stars with rotation periods at very low $L_X$ because
the amplitude of
photometric variability in the optical becomes diminishingly small,
ultimately smaller than the minimum signal detectable ($\Delta I \sim 0.03$ mag)
by existing rotation-period studies of the ONC \citep{stass99,herbst02}.

In light of the fact that stars with rotation periods have high
X-ray luminosities, it is perhaps not surprising that these stars 
appear to be in the super-saturated regime.
But if these stars are indeed super-saturated as
Fig.\ \ref{fig-lxvsrossby} implies,
then the optical variability data
would seem to imply a qualitatively different picture for the
surfaces of super-saturated stars than that commonly assumed.
The mental image often invoked in the context of saturation is that
of a star whose surface has become completely threaded by magnetic flux 
tubes, resulting in spot coverage fractions approaching unity. Yet in
Fig.\ \ref{lx-spots} there are
stars at both low and high $L_X/L_{bol}$ that show relatively small
amplitudes of optical variability, suggesting that
spot coverage among many of these ``super-saturated" stars is relatively light.

On the other hand, Fig.\ \ref{lx-spots} may be telling us that
these stars do indeed have spots covering large fractions of their surfaces,
but that we are seeing changes in the magnetic topologies of these stars
as a function of $L_X/L_{bol}$.
For example, stars at lower $L_X/L_{bol}$ may represent stars with relatively
disorganized surface fields that produce 
relatively small spots more-or-less uniformly distributed on
the stellar surface. Such small, uniformly distributed spots would 
produce only low-amplitude variability 
in the optical even if they cover a large fraction of the stellar surface. 
In contrast, stars with larger $L_X/L_{bol}$ could represent cases
where the magnetic field has become more coherently organized into
relatively large spots that are distributed more asymmetrically on
the stellar surface, thereby giving rise to larger photometric variability
in the optical. 
That not all stars with large $L_X/L_{bol}$ have correspondingly large $\Delta I$ 
is perhaps simply due to geometrical effects (varying spot sizes/temperatures, 
spot latitudes,
inclination angles, etc.), or it may suggest that strong magnetic fields
do not instantaneously arrange into organized configurations.

This interpretation is similar to that proposed by \citet{barnes03b},
who argues that stars in the super-saturated regime are
cases in which the stellar magnetic field has not yet become sufficiently
organized to couple the stellar interior to the surface, and therefore
the star's rotation is not effectively braked. \citet{barnes03b} further
argues that as the stellar magnetic field becomes more organized
and achieves maximum strength, it becomes more deeply rooted, the
X-ray luminosity also reaches maximum strength (saturation), and magnetic 
braking begins to affect the entire star. 
In this way, \citet{barnes03b} offers a possible explanation for the positive
correlation observed between $P_{rot}$ and $L_X/L_{bol}$ among stars in the
super-saturated regime.
These are speculative ideas to be sure; our aim here is to provide
additional observational fodder to the question of what super-saturation 
is really telling us about the magnetic nature of PMS stars.

At any rate, if we accept the inference that stars in the ONC with rotation 
periods do represent the super-saturated regime of the 
rotation/X-ray relationship, then the question arises
whether there is evidence for an unseen linear regime in the 
rotation/X-ray relationship. Fig.\ \ref{fig-lxvsrossby} tells us 
that there are indeed stars with sufficiently low $L_X / L_{bol}$, 
but do these stars also rotate more slowly? While the available
$v\sin i$ data do not show a one-to-one relationship between $v\sin i$ 
and $L_X$, we do find evidence that slower rotators do indeed have lower $L_X$
(\S \ref{rotation}), hinting at behavior qualitatively consistent with the 
linear regime of the rotation/X-ray relationship. 

Thus, a picture begins to emerge from the data in which 
X-ray luminosity does appear to be related to stellar rotation among PMS 
stars in the ONC. Stars with rotation periods, biased as they are in
$L_X$, may represent the super-saturated and saturated regimes, and some 
stars lacking rotation periods
may represent the saturated and (at least part of) the linear regime,
implying a population of very slow rotators among these stars. 

An alternative to the slow-rotator explanation for the lower $L_X$ of
stars without rotation periods is that stars without rotation periods
are predominantly active accretors, and that it is accretion that is acting
to suppress the $L_X$ of these stars (see \S \ref{acc-lum}). Indeed, among
the sample of stars from \citet{feig02} that lack rotation periods, those
with spectroscopic signatures of active accretion (i.e.\ EW(\ion{Ca}{2}) 
$\le -1$ \AA) outnumber those without such signatures by 2:1. To examine
this possibility more fully, we have compared the hardness ratios (HRs) of stars
with and without rotation periods, since HR is also correlated with accretion 
(accretors produce harder HRs; see \S \ref{acc-hardness}). We find that
the HRs of stars without rotation periods are marginally harder than those
with rotation periods; a K-S test yields a probability of 1\% that the 
distributions of HRs for the two groups are the same. Compared to the
result in \S \ref{acc-hardness}---where we found a highly statistically 
significant difference in HR for accretors vs.\ non-accretors---this suggests
that, for the particular mix of stellar masses and accretion properties in
the non--$P_{\rm rot}$ sample, accretion is only weakly related to the 
lower average $L_X$ of these stars. The significance of the effect is, 
nonetheless, comparable to the $v\sin i$ effect described above.

Discerning whether, or to what extent, 
the lower average $L_X$ of stars lacking rotation periods is due to
accretion or slower rotation
remains an open observational question. Unfortunately, the existing
$v\sin i$ study of \citet{rhode} did not have sufficiently high spectral
resolution to place stringent lower limits on the rotation rates of these 
stars. It would thus be valuable to have high-resolution $v\sin i$ measurements 
targeting stars with very low $L_X$ and lacking $P_{rot}$ in order 
to better constrain the slow
extremes of rotation among stars that may represent the saturated and linear
regimes ($P_{\rm rot} \gtrsim 20$ days) of the rotation/X-ray relationship. 

Finally, we call attention to the fact that stars with rotation periods,
despite evincing stable optical photometric variability with low levels
of stochasticity (else their rotation periods would be difficult to 
measure), nonetheless show elevated levels of variability in
X-rays (\S \ref{var}). This may suggest that the mechanism(s) 
responsible for X-ray variability are decoupled from the mechanism(s) often 
attributed to stochastic optical variability in PMS stars (i.e.\ accretion), 
as we now discuss.

\subsection{Accretion\label{disc-accretion}}
It is now generally accepted that most, if not all, PMS stars undergo a
phase of active accretion whereby circumstellar material, perhaps channeled
by stellar magnetic field lines, is deposited onto the stellar surface. 
Models of this accretion process \citep{calvet,gullbring,valenti} have had 
some success in explaining the continuum excesses often observed in the UV
among PMS stars as being due to the energetic shock that arises when accreted
material impacts the stellar surface. Accretion is also typically
implicated as the source of the stochastic, optical variability that is
a defining characteristic of classical T~Tauri stars (CTTS) \citep{herbst94}.
It is appropriate to ask, therefore, whether X-rays from PMS stars may also
have their origins, at least partly, in accretion.

We have already seen that X-ray variability is ubiquitous among the PMS
stars in this study, despite the fact that the majority of these stars
are weak-lined T~Tauri stars (WTTS), as they do
not show spectroscopic indicators of active accretion (\S \ref{acc-flaring}).
But perhaps accretion acts nonetheless to noticeably affect the
X-ray emission of these stars.
Indeed, we have seen that accretors and non-accretors do differ both in their
X-ray luminosities (\S \ref{acc-lum}) and X-ray hardness (\S \ref{acc-hardness}).
Here we investigate these differences in greater detail. 

We begin by reviewing the evidence, both from this study and from others
in the literature, for a difference in the X-ray luminosities
between accretors and non-accretors. We then present a simple model that
explains these differences naturally in terms of enhanced X-ray 
{\it absorption} among stars with active accretion, due to the presence 
of magnetospheric accretion columns.

\subsubsection{Differences in X-ray luminosities between accretors
and non-accretors\label{disc-lx-acc}}
Among PMS stars in a variety of star formation regions, there 
appears to be strong evidence for a difference in X-ray luminosity
between accretors and non-accretors, in the sense that accretors tend to 
be underluminous in X-rays relative to non-accretors. A summary of the 
situation with a re-analysis of {\it ROSAT\/} data is presented in 
\citet{flac-acc} for the ONC, NGC 2264, and Chameleon I. 
Similar results are found by \citet{neuhauser95} in Taurus-Auriga.

However, the most recent observations in Orion present two different results. 
\citet{flac-basic} find that that the difference in the 
median $L_X$ between accreting and non-accreting stars is about one order 
of magnitude in the 0.25--2 M$_\odot$ range, in agreement with the earlier
{\it ROSAT\/} findings. These authors use the EW of the \ion{Ca}{2}
lines, as reported by \citet{hill98}, to distinguish accretors from non-accretors.
Their study is based on a single exposure with {\it Chandra/HRC} (30' by 30') 
centered on $\theta^1$Ori C. 
Optical observations catalog 696 cluster 
members in the field, 342 of which are detected in the HRC image. 
Of the 696 possible members, a subset (304 stars) have EW(\ion{Ca}{2}) 
$< -1$ (108 X-ray detected, 58 undetected) or EW(\ion{Ca}{2})
$> 1$ (54 X-ray detected, 84 undetected). 
As the HRC instrument does not provide
spectral information, the authors assume a 
fixed plasma temperature for all sources and gas column density proportional 
to optical extinction in order to derive X-ray luminosities.

In contrast, \citet{feig02} find no 
difference in the distributions of CTTS and WTTS with respect to X-ray 
luminosity. Here, the distinction between CTTS and WTTS is made in terms of 
K-band excess, which is taken to indicate the presence of an accretion disk. 
Their study is based on the same {\it Chandra/ACIS} observations that we
use in our own analysis.
The ACIS image (17' by 17') is centered 22'' west of $\theta^1$Ori C. 
In that region there are 529 optically detected stars, 525 of which are 
detected in the ACIS exposure. 

The discrepancy between the findings of \citet{flac-acc} and \citet{feig02} 
can be resolved by noting 
that while infrared indicators signal the presence of a disk, this does
not necessarily signal the presence of active accretion: the presence of
a disk is presumably a prerequisite for accretion to occur, but not
necessarily vice-versa.
Indeed, using the same spectroscopic proxy for accretion as \citet{flac-acc},
our analysis above (\S \ref{acc-lum}) confirms the findings of \citet{flac-acc}
within the same ACIS observations used by \citet{feig02}.

We thus take the finding of a difference in X-ray luminosity between 
accretors and non-accretors, as shown in Fig.\ \ref{lx-acc}, to be secure.
In addition, we have found evidence for a difference between accretors and
non-accretors in terms of X-ray hardness (Fig.\ \ref{hr-acc}).
We now proceed to examine possible explanations for these differences.

\subsubsection{Explanation: Enhanced X-ray emission or circumstellar
absorption?\label{disc-explanation}}
PMS stars undergoing active accretion show systematically lower
X-ray luminosities and harder X-ray hardness ratios (HR) than their non-accreting
counterparts. This suggests that either: (a) the X-ray emission from accretors 
is intrinsically different in its spectral properties, namely, more
concentrated to higher X-ray energies (i.e.\ harder); or (b) the X-ray emission 
from the accretors is intrinsically similar to that from non-accretors, but has 
been processed by circumstellar gas, preferentially attenuating X-rays at softer
energies.

In the magnetospheric picture of accretion, CTTS are encaged in funnels of
inflowing gas \citep{muzerolle} with densities ranging from $10^{12}$ to 
$10^{14}$ cm$^{-3}$ \citep{calvet}. These funnels may be 
0.1 R$_\odot$ thick, which implies that hydrogen column densities larger than 
$10^{22}$ cm$^{-2}$ are possible. The exact amount of gas column 
will depend on the accretion rate and on the detailed geometry of the 
accretion flows but, as we show below, this amount of hydrogen column is 
potentially sufficient
to both attenuate and harden the X-rays observed from CTTS.

To investigate this further, we first need to obtain 
the intrinsic (corrected for ISM absorption) X-ray characteristics of the
{\it Chandra} sources. 
The X-ray luminosities and HRs that we have so far used in our analysis
have not been corrected for the attenuation and hardening caused by absorption
due to interstellar gas. In some star formation regions, this is an important 
issue. For example, \citet{neuhauser95} have shown that in Taurus 
the reddening toward CTTS is significantly higher than toward WTTS, which
could produce systematic differences in $L_X$ and HR similar to what we
have observed.
In the \citet{feig02} data there is no evidence for a systematic difference 
in extinction between 
accretors and non-accretors; the extinction properties of both groups are the 
same to within 20\%. Nonetheless, there may still be 
individual differences in extinction that could act to alter 
the medians in Figs.\ \ref{lx-acc} and \ref{hr-acc}. 

\begin{figure}
\plotone{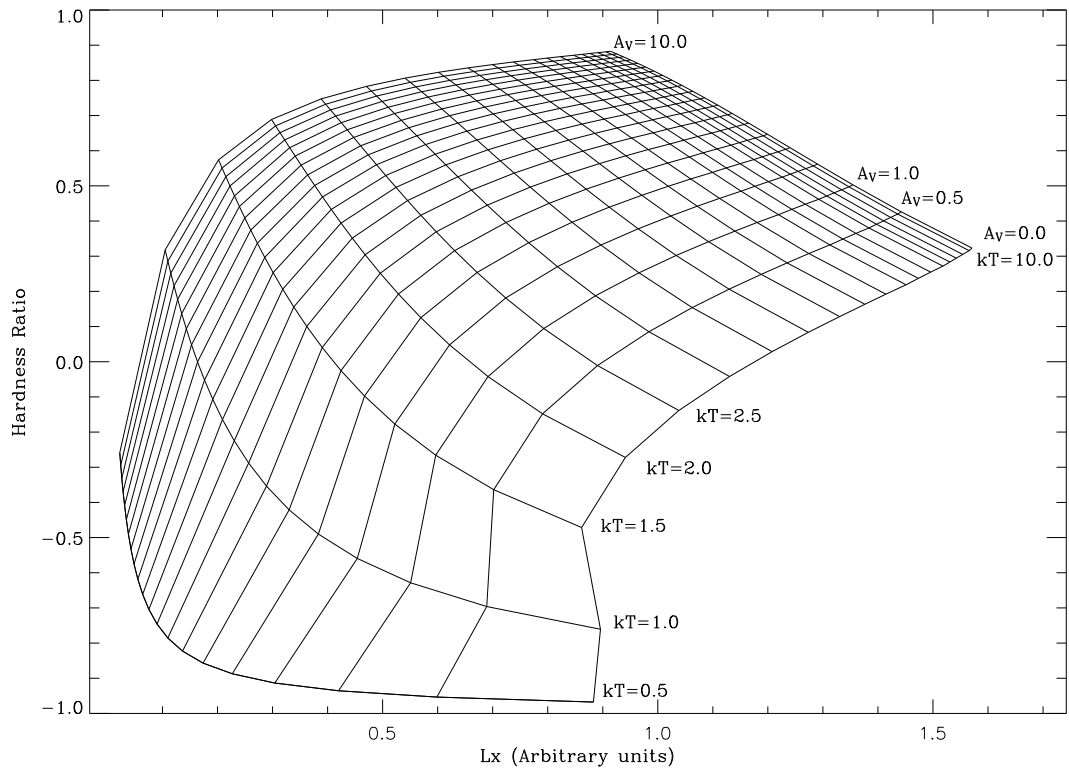}
\caption{\label{model} Theoretical hardness ratios as a function of Lx for 
different values of $kT$ and $A_V$. The curves are marked with optical 
extinction and $kT$ values. The abcisa values are arbitrary up to a 
multiplicative constant.}
\end{figure}

\begin{figure}
\epsscale{0.8}
\plotone{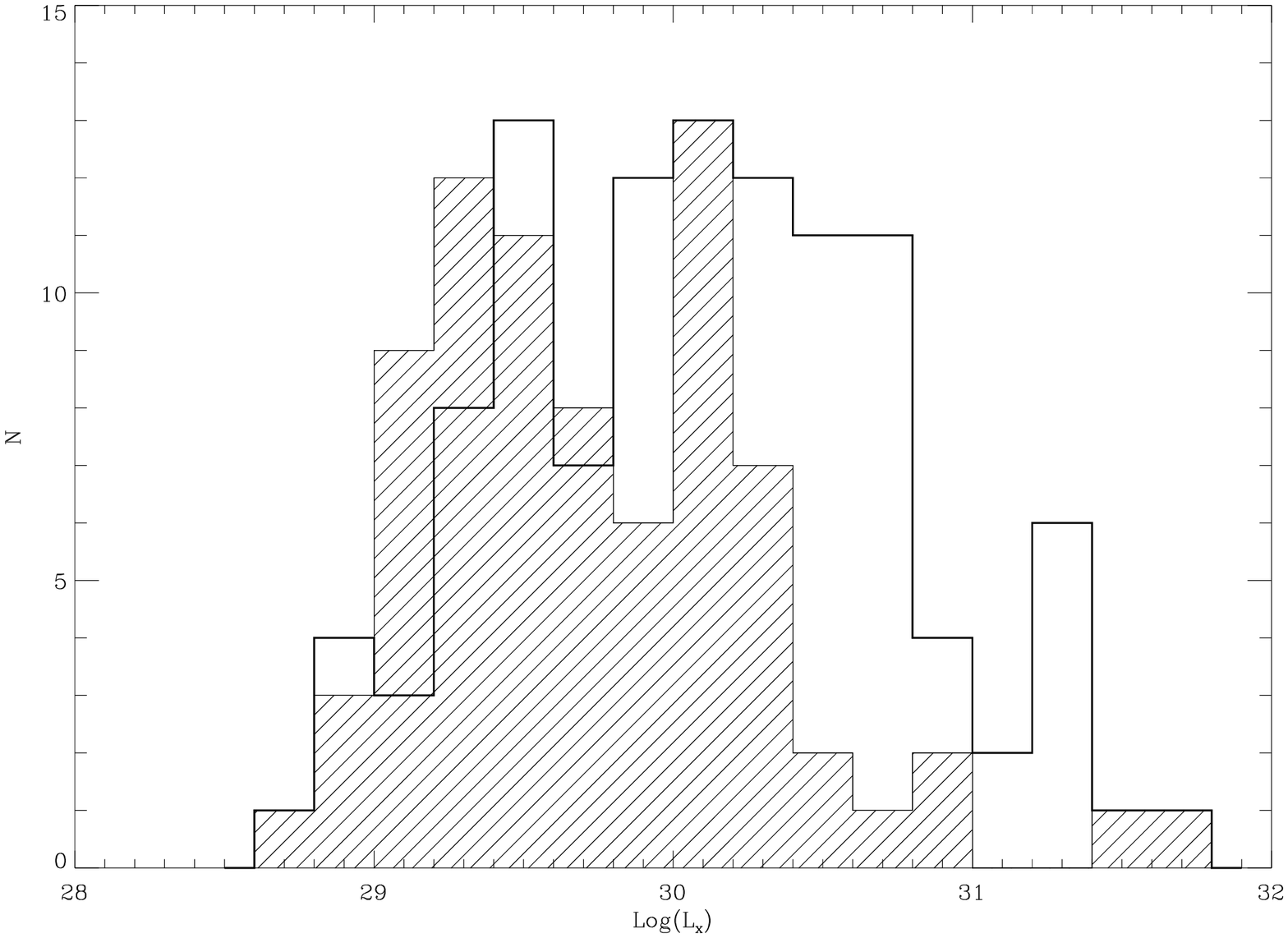}
\plotone{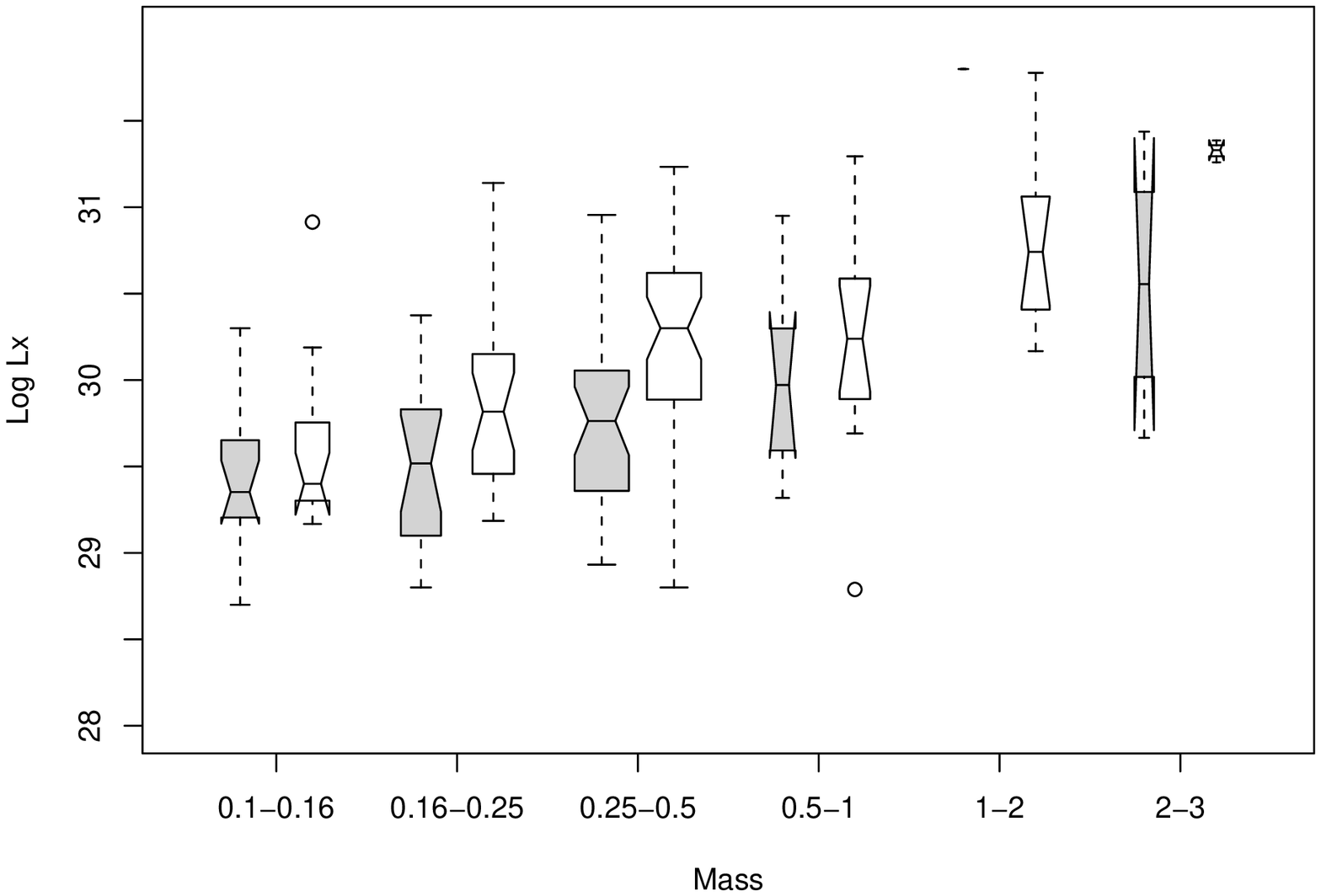}
\caption{\label{corr_boxes}X-Ray luminosities, corrected for interstellar 
extinction. See Fig.\ \ref{lx-acc} for an explanation of symbols. 
A K-S test indicates that the probability of the two histograms being 
drawn from the same parent distribution is $4\times 10^{-4}$.}
\end{figure}

\begin{figure}
\plotone{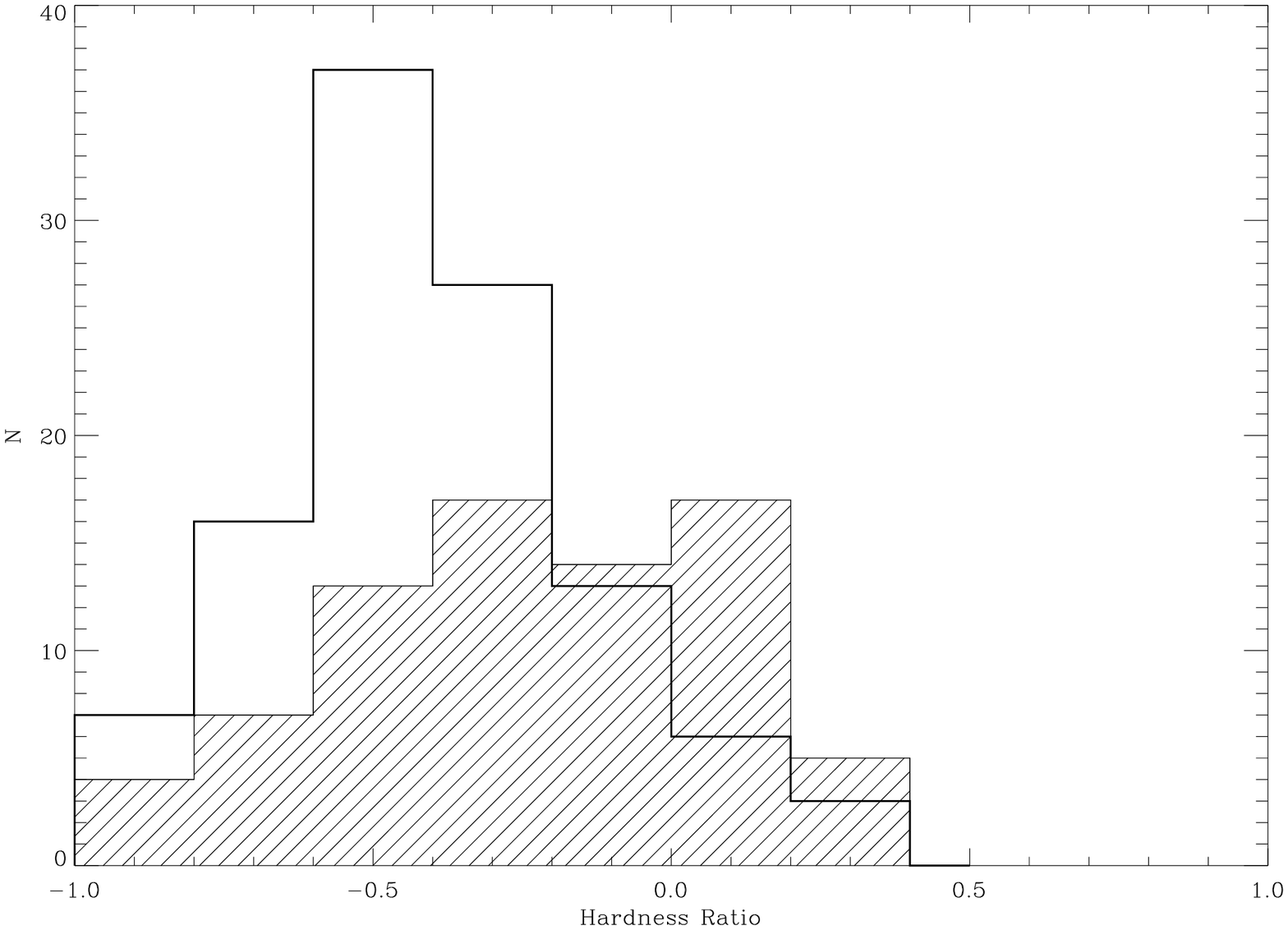}
\plotone{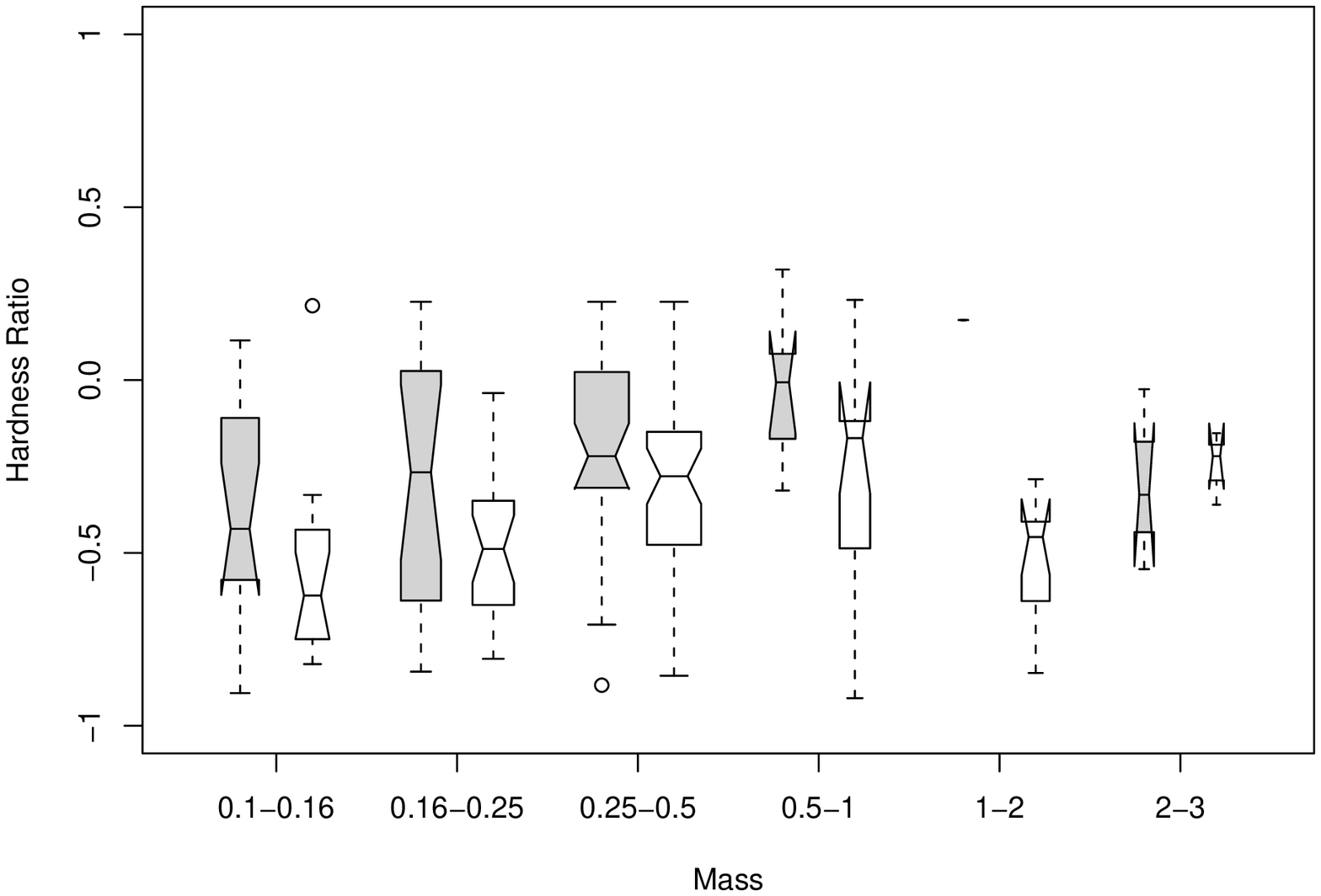}
\caption{\label{corr_hard}HR values, corrected for interstellar extinction. See 
Fig.\ \ref{hr-acc} for an explanation of symbols. A K-S test indicates that 
the probability of the two histograms being drawn from the same
parent distribution is $2\times 10^{-4}$.}
\end{figure}

In order to correct for interstellar reddening we have performed 
the following analysis. 
We first calculate HR and $L_X$ values for a grid of hydrogen 
column densities and plasma temperatures (Fig.\ \ref{model}). To generate these 
models, we used the Xspec code \citep{arnaud}, 
version 11.2, assuming a uniform plasma with $0.3\times$ solar elemental 
abundances. As in \citet{feig02}, continuum and line emission strengths were 
evaluated using the MEKAL code \citep{mewe}, and  
X-ray absorption was modeled using the cross sections of
\citet{morrison}. For each star in the \citet{feig02} database, we take
the HR and $L_X$ values reported by them
and extinctions ($A_V$) from \citet{hill97}. We then use the
relation $N_H=2\times 10^{21}A_V$ to convert the observed extinctions into
a measure of the hydrogen column density toward each star. 
From Fig.\ \ref{model}, we obtain
the ratio between the observed luminosity and the 
luminosity corrected for reddening. For example, if a star is observed to have
HR $=0.0$ and $A_V=1.5$, Fig.\ \ref{model} tells us that the observed
$L_X =0.6$ (arbitrary units) and that the intrinsic 
$L_X = 0.9$ (obtained by moving in constant $kT$ to $A_V=0$),
implying that the X-ray luminosity has been extincted by a factor of $\sim 0.7$
and that the true HR is $\sim -0.3$. In this way we obtain 
corrected values of $L_X$ and HR for each star. The results are shown 
in Figs.\ \ref{corr_boxes} and \ref{corr_hard}. 
The temperature obtained by this procedure should be regarded as an 
``effective'' plasma temperature, as individual fits suggest that in some 
cases multiple plasmas, each with a different temperature, are necessary to 
reproduce the observations. The procedure also assumes that the plasma 
is in ionization equilibrium
\citep{ardila}.

After correcting for reddening, the differences in the histograms persist
(Figs.\ \ref{corr_boxes}a and \ref{corr_hard}a),
although when plotted as functions of mass 
(Figs.\ \ref{corr_boxes}b and \ref{corr_hard}b)
the differences between the accretors and non-accretors become more subtle. 
It is therefore legitimate to ask whether the differences 
in the histograms are real, considering the dependence of
$L_X$ and HR on mass. For example, the presence of proportionately more 
non-accretors than accretors at higher masses could potentially explain the  
differences in the histograms. A two-way analysis of variance indicates that 
the $L_X$ averages of accretors and non-accretors, after 
eliminating the effect of the mass, have a probability of 
$1 \times 10^{-3}$ of being the same. For HR, the probability is 
$1\times 10^{-4}$. 
In other words, there is a statistically significant difference 
between accretors and non-accretors, both in HR and in $L_X$, 
even after controlling for the mass dependence.  Interestingly, 
HR appears to increase (albeit weakly) with mass---the analysis of 
variance indicates that the probability of all the means in the mass 
bins being the same is $10^{-3}$---perhaps implying that more 
massive stars have hotter chromospheres. 
This is not due to the fact that higher mass stars have higher $L_X$; 
HR is scale-independent, so overall increases in $L_X$ do not affect it. 

Differences in HR values between CTTS and WTTS have been 
reported in the literature 
for Taurus, Lupus, Chameleon, Sco-Cen, and the TW Hya association
\citep{neuhauser94,krautter,neuhauser95,kastner}. 
All these are based on {\it ROSAT} data, for which two different hardness 
ratios are traditionally defined in the literature: 
HR1 $=(Z_{h1}+Z_{h2}-Z_s)/(Z_{h1}+Z_{h2}+Z_s)$---where $Z_{h1}$ is 
the count rate from 0.5 to 0.9 keV, $Z_{h2}$ is from 0.9 to 2 keV, 
and $Z_s$ is from 0.1 to 0.4 keV---and HR2 $=(Z_{h1}-Z_{h2})/(Z_{h1}+Z_{h2})$. 
Note that the two ``hard" {\it ROSAT} bands are equivalent to the ``soft" 
{\it Chandra} band so the results from {\it Chandra} and {\it ROSAT} are not 
directly comparable. In the {\it ROSAT} observations, and for these star formation 
regions, the WTTS are as a group significantly softer than the 
CTTS in the HR1 ratio, while the two populations have similar HR2 ratios. 
Our analysis shows that the difference reappears in the higher energy 
{\it Chandra} HR ratio, which samples energies up to 8 keV. 
\citet{neuhauser95}, finding no difference in 
emission temperatures between CTTS and WTTS in Taurus, and
considering different star-forming regions with different 
extinction characteristics, argue that this difference in HR is due to 
absorption in the circumstellar environs of the CTTS
(circumstellar disks, remnant nebulae and envelopes, outflows, etc.).

For the observations presented here, the differences in $L_X$ and HR between 
accretors and non-accretors are consistent with a picture in which
CTTS have intrinsically similar X-ray emission properties as WTTS, 
with X-rays from the former being extincted by circumstellar gas in
amounts consistent with that predicted for magnetospheric accretion columns. 
The median HR (corrected for absorption) of the non-accretors in our sample 
is $-0.40$ with $\sigma=0.3$. For the accretors, 
the value is $-0.23$ with $\sigma=0.3$. Assuming that the difference is due 
to gas absorption, we can use Fig.\ \ref{model} to obtain the gas column density. 
If the mean HR of the non-accretors in our sample ($-0.40$) represents the 
intrinsic HR of a T Tauri star, this implies 
(following the $A_V=0.0$ curve) a plasma temperature of $kT\sim1.7$ keV. 
The curves are marked in dust extinction 
magnitudes, but in this excercise we are using them to correct for gas absorption 
only. If we follow the line of constant $kT$ to higher hardness ratios, 
we reach HR $\approx -0.2$ at $A_V\approx 1.0$, which implies $2\times 10^{21}$ 
cm$^{-2}$. 
In this case, the ratio in $L_X$ between accretors and non-accretors would be 
$\sim0.7$. Given the width of the HR histograms, column densities as large as  
$10^{22}$ cm$^{-2}$ of gas may be necessary. These produce ratios in 
$L_X$ as large as 0.8 dex, which is consistent with Fig.\ \ref{corr_boxes} 
and with the results of \citet{flac-acc}.

On the other hand, \citet{kastner} argue, on the basis of {\it Chandra} X-ray 
spectroscopy, that the X-ray emission from TW Hya is due to the accretion shock 
at the base of the accretion column,
and  not simply to attenuated WTTS emission. The differential emission meassure 
is quite unlike that of other active evolved stars (even though it is not clear 
what one should expect for a PMS star). TW Hya is a 10 Myr old, 
0.7 M$_\odot$ PMS star with $L_X\sim10^{30}$ erg/sec, and so it has a very 
average position in our $L_X$ vs.\ mass diagram.  This star poses a puzzle for 
the arguments presented here in favor of a common origin for X-ray emission 
in CTTS and WTTS. Its accretion rate has been reported as being 5--100 
$\ 10^{-10}$ M$_\odot$/yr \citep{muzerolle,alencar}, and if the lower limit is 
right, one would expect essentially no gas attenuation. In addition, 
coronal activity decreases with age, and so perhaps the observations of TW Hya 
are not applicable to younger samples. Certainly, X-ray spectroscopic 
observations of young WTTS and CTTS are needed before this issue can be 
fully resolved.

\section{Summary and Conclusions\label{summary}}
We have re-analyzed all archival {\it Chandra/ACIS} observations of 
pre--main-sequence (PMS) stars with optically determined rotation periods
in the Orion Nebula Cluster (ONC). Our aim is to investigate
the relationship between 
X-rays and the physical mechanisms most likely related to their
production in PMS stars: rotation and accretion.
Our analysis procedures include filtering
of flare events in the X-ray data in an attempt to determine X-ray 
luminosities that are free of the stochasticity introduced by such
events.

The primary findings of this study are as follows:
\begin{enumerate}
\item 
Stars with optically determined rotation periods are 
more X-ray luminous, and are more likely to be X-ray 
variable, than are stars without optical rotation periods.
We show that the bias to high $L_X$ is not due to a magnitude bias in 
optical rotation-period studies of the ONC;
rather, it is due to the diminishingly small amplitude of optical
variability among stars with smaller $L_X$, precluding detection of
their rotation periods.

\item
Stars with optically determined rotation periods have a mean $L_X/L_{bol}$ 
near, but lower than, the ``saturation" value of $\sim 10^{-3}$, implying that 
these stars are in the saturated or super-saturated regimes of the
X-ray/rotation relationship, consistent with their Rossby numbers.
There is a marginally significant ($\sim 3\sigma$) correlation between 
$L_X/L_{bol}$ and $P_{\rm rot}$,
with the more rapidly rotating stars showing lower $L_X/L_{bol}$, as is
seen among super-saturated stars on the main sequence.

\item
Compared to these stars,
stars {\it without} rotation periods show a larger
range of $L_X/L_{bol}$---comparable, in fact, to that found among
main sequence stars. We consider the possibility that, among these, some stars
may lie at the beginnings of the ``linear" regime of the X-ray/rotation
relationship. Using $v\sin i$ data from the literature we find that,
among these stars lacking known rotation periods,
slower rotators do indeed show lower X-ray luminosities than 
do rapid rotators. This relationship is not one-to-one, however.
It is also possible that the lower $L_X$
among stars lacking rotation periods is instead due to
the higher incidence of active accretion 
among these stars, a possibility for which we also find weak evidence. 
The statistical significance of these two effects---$v\sin i$ and 
accretion---are comparable.
Measurements of $v\sin i$ sensitive to very slow rotators ($\lesssim 5$ km/s)
would be of great value in furthering our
understanding of X-ray production at the slow extremes of PMS rotation.
PMS stars in the linear regime should have $P_{\rm rot} \gtrsim 100$ days,
assuming a typical convective turnover timescale of $\tau_c \sim 800$ days.
Such long rotation periods have yet to be observed among PMS stars.

\item
Stars in the ONC with spectroscopic signatures of active accretion show
significantly harder X-ray spectra and lower X-ray luminosities than 
their non-accreting counterparts.
These observations can be explained quantitatively by a
model in which accretors and non-accretors have intrinsically similar
X-ray emission properties, with the differences in $L_X$ and hardness
ratio being due to absorption of soft X-rays by magnetospheric accretion 
columns.
\end{enumerate}

Taken together, these findings hint that there in fact exists a
rotation-activity relationship among PMS stars in the ONC, and
suggest that rotation---not accretion---is the primary driver of X-ray 
emission in low-mass ($M \lesssim 3$ M$_\odot$) PMS stars at 1 Myr.
Indeed, our finding that stars with rotation periods show
elevated levels of X-ray variability, despite showing little stochastic 
variability in the optical, further implies that X-ray variability has
its origins in processes that are more or less independent of the
processes responsible for stochastic variability in the optical (i.e.\
accretion).

Finally, our findings raise questions about the true physical meaning of
``saturation" in PMS stars.
It is intriguing that stars with optically determined rotation periods
all appear to lie in the super-saturated regime
yet show diminishingly small amplitudes of optical
variability at low $L_X$.  It is possible that
spots on the surfaces of these stars become non-existent below a certain
$L_X$ threshold. 
On the other hand, we speculate that
the low amplitude of optical variability may be due to magnetic
topologies in which the stellar surface is indeed largely covered by
spots, but spots 
that are more-or-less randomly
distributed over the stellar surface, thereby producing only very small 
photometric signals in the optical. 
More organized magnetic topologies may be present in stars with higher
$L_X$, such that larger spots asymmetrically distributed on the stellar
surface are possible.
In this picture, these latter stars might be those
whose global fields have become sufficiently organized and deeply rooted 
so as to begin effecting magnetic braking of the stellar rotation, a picture
similar to that recently put forward by \citet{barnes03a,barnes03b}.

\acknowledgments
We acknowledge funding under Chandra Award
Number AR2-3001X issued by the Chandra X-Ray Observatory Center, which
is operated by the Smithsonian Astrophysical Observatory on behalf
of NASA under contract NAS8-390073.
We also gratefully acknowledge the useful comments of the anonymous referee.

\clearpage

\clearpage

\begin{deluxetable}{rrrrrr}
\tablecolumns{6}
\tablewidth{0pt}
\tabletypesize{\scriptsize}
\tablecaption{Study sample\label{table-sample}}
\tablehead{
\colhead{ID\tablenotemark{a}} & \colhead{$P_{rot}$} & \colhead{$M_\star$} & 
\colhead{$\log L_{bol}/L_\odot$} & \colhead{$\Delta (I-K)$} & \colhead{EW(\ion{Ca}{2})} \\
\colhead{} & \colhead{days} & \colhead{M$_\odot$} & \colhead{} & \colhead{mag} &
\colhead{\AA}
}
\startdata
106 & 1.70 & 0.21 & $-$0.29 & 0.10 & 1.5 \\ 
111 & 4.94 & 0.42 & $-$0.15 & 0.24 & 2.2 \\ 
116 & 2.34 & 0.69 & 0.20 & 0.09 & 1.6 \\ 
118 & 1.07 & 0.13 & $-$0.61 & $-$0.32 & 0.0 \\ 
122 & 0.98 & 0.14 & $-$0.54 & $-$0.17 & 0.0 \\ 
123 & 6.63 & 1.37 & 0.28 & 1.28 & 0.0 \\ 
128 & 8.83 & 0.15 & 0.28 & 0.32 & 0.0 \\ 
133 & 2.03 & 0.29 & $-$0.25 & 0.26 & 1.6 \\ 
136 & 8.65 & 0.28 & 0.06 & \nodata & \nodata \\ 
140 & 4.58 & 0.17 & $-$0.19 & $-$0.29 & 3.8 \\ 
\enddata
\tablenotetext{a}{Designation from \citet{hill97}.}
\tablecomments{Table 1 is availble in its entirety in the electronic
edition of the {\it Astronomical Journal}. A portion is shown here for
guidance regarding its form and content.}
\end{deluxetable}

\clearpage

\begin{deluxetable}{rcrrcc}
\tablecolumns{6}
\tablewidth{0pt}
\tabletypesize{\scriptsize}
\tablecaption{X-ray properties of study sample\label{table-lx}}
\tablehead{
\colhead{ID\tablenotemark{a}} & \colhead{Exposure\tablenotemark{b}} & 
\colhead{$\log L_X$\tablenotemark{c}} & 
\colhead{$\log (L_X)_F$\tablenotemark{d}} & 
\colhead{Variability\tablenotemark{e}} & \colhead{Flag\tablenotemark{f}} \\
\colhead{} & \colhead{} & \colhead{erg/s} & \colhead{erg/s} & \colhead{} &
\colhead{}
}
\startdata
174 & G2 & 29.8 & 29.4 & Const & 0 \\ 
175 & G1 & 29.4 & 29.8 & LTVar & 0 \\ 
175 & G2 & 30.3 & 29.8 & LTVar & 1 \\ 
177 & G2 & 30.0 & 30.2 & Flare & 0 \\ 
177 & G1 & 30.3 & 30.2 & Flare & 1 \\ 
178 & T & 30.1 & \nodata & PosFl & 1 \\ 
187 & G2 & 30.5 & 30.3 & Flare & 1 \\ 
187 & G1 & 30.4 & 30.3 & Flare & 0 \\ 
188 & G2 & 29.5 & 29.7 & PosFl & 0 \\ 
188 & G1 & 30.6 & 29.7 & PosFl & 1 \\ 
\enddata
\tablenotetext{a}{Designation from \citet{hill97}.}
\tablenotetext{b}{Source of measurement. G1: First Garmire exposure;
G2: Second Garmire exposure; T: Tsujimoto exposure.}
\tablenotetext{c}{X-ray luminosity from this study.}
\tablenotetext{d}{X-ray luminosity from \citet{feig02}.}
\tablenotetext{e}{X-ray variability, from \citet{feig02} or from this
study if source not included in \citet{feig02} study. `Const' indicates
a non-variable light curve, `Flare' indicates a light curve with a
clear flare, and 'PosFl' indicates a light curve that possibly includes
a flare.}
\tablenotetext{f}{Quality flag (see text). Measurements with a `1' 
are those used in our analysis.}
\tablecomments{Table 2 is availble in its entirety in the electronic
edition of the {\it Astronomical Journal}. A portion is shown here for
guidance regarding its form and content.}
\end{deluxetable}

\end{document}